\documentclass[twocolumn,prd,showpacs,floatfix]{revtex4}
\usepackage{graphicx}
\usepackage{dcolumn}
\usepackage{bm}
\begin{document}
\title{ Boost-invariant Hamiltonian approach to heavy quarkonia }
\author{ Stanis{\l}aw D. G{\l}azek } 
\author{ Jaros{\l}aw M{\l}ynik}
\affiliation{ Institute of Theoretical Physics, Warsaw University, 
              ul.  Ho{\.z}a 69, 00-681 Warsaw, Poland }
\date{10 October 2006}
\begin{abstract}
Light-front Hamiltonian formulation of QCD with only one
flavor of quarks is used in its simplest approximate version
to calculate masses and boost-invariant wave functions of $c
\bar c$ or $b \bar b$ mesons. The quark-antiquark
Hamiltonian is obtained in the lowest (second) order of a
weak-coupling expansion scheme for Hamiltonians of effective
particles in quantum field theory. The derivation involves a
heuristic ansatz for a gluon mass-gap that is meant to
account for non-Abelian color dynamics of virtual effective
particles in the Fock components with one or more effective
gluons and can be improved order-by-order in the expansion.
Fortunately, the resulting quark-antiquark Hamiltonian does
not depend on any details of the ansatz within a large class
of possibilities. It is shown that in the Hamiltonian
approach in its simplest version the strong coupling
constant $\alpha$ and quark mass $m$ (for suitable values of
the renormalization group parameter $\lambda$ that is used
in the calculation), can be adjusted so that a) masses of 12
lightest well-established $b \bar b$ mesons are reproduced
with accuracy better than 0.5\% for all of them, which means
50 MeV in a few worst cases and on the order of 10 MeV in
other cases, or b) masses of 11 lightest $c \bar c$ mesons
are reproduced with accuracy better than 3\% for all of
them, which means better than 100 MeV in a few worst cases
and on the order of 10 MeV in the other cases, while the
parameters $\alpha$ and $m$ are near the values expected in
the cases a) and b) by analogy with other approaches. A
4th-order study in the same Hamiltonian scheme will be
required to explicitly include renormalization group running
of the parameters $\alpha$ and $m$ from the scale set by
masses of bosons $W$ and $Z$ down to the values of $\lambda$
that are suitable in the bound-state calculations. In
principle, one can use the Hamiltonian approach to describe
the structure, decay, production, and scattering of heavy
quarkonia in all kinds of motion, including velocities
arbitrarily close to the speed of light. This work is
devoted exclusively to a pilot study of masses of the
quarkonia in the simplest version of the approach.
\end{abstract}
\pacs{12.38.Aw, 11.15.Tk, 12.38.Lg}
\maketitle
\section{ Introduction }
\label{sec:I}
In the Hamiltonian approach to QCD that is employed here,
the calculation of masses of heavy quarkonia does not
involve the well-known notions of scattering states for
quarks and gluons, Feynman diagrams, path integral,
euclidicity postulate, lattice gauge theory, or non-trivial
vacuum expectation values. Instead, a renormalization group
procedure for effective particles in quantum field theory
(RGPEP, see below) is applied to quarks and gluons in
canonical light-front (LF) QCD and leads to an effective
boost-invariant Hamiltonian whose eigenvalue problem is
expected to provide a first approximation to the true
dynamics of the theory. The eigenvalues of the Hamiltonian
are equal to the masses of the quarkonia (actually, squares
of the masses) instead of their energies in a specific frame
of reference. The approximate dynamical picture studied here
can be valid only for a set of states near the low end of
the mass spectrum. 

The LF formalism is developed in a Fock space. Creation
and annihilation operators for effective quarks and
gluons are calculable in a perturbative expansion using
RGPEP (a brief review of the method is provided in the
next section) and the basis states in the Fock space
are formed by acting with the calculable creation
operators on the state of vacuum. The effective
Hamiltonian eigenvalue problem exists in the momentum
representation and would not be local if one attempted
to write it in a position space. Nevertheless, the
simplest approximate momentum-space dynamics respects
not only the boost symmetry but also the rotational
symmetry. Therefore, one can describe the structure of
lowest-mass eigenstates corresponding to well-known $b
\bar b$ and $c \bar c$ mesons using the spectroscopic
scheme that is quite analogous to the one adopted in
non-relativistic quantum mechanics with potential
forces that respect rotational symmetry in a meson
center-of-mass frame of reference. Thus, the LF
approach has a potential to be helpful in solving
conceptual problems with Poincar\'e symmetry in quantum
theories \cite{Dirac1,Dirac2} and the results described
here can be considered an indicator of existence of a
reasonable candidate for a new expansion method for
solving theories as complex as QCD \cite{Wilson2004}.
However, the approach is still in its infancy. The
crude, heuristic study described here is merely a small
step that needs to be taken on the way to find out if
the RGPEP can work for heavy quarkonia in LFQCD as
outlined in \cite{ho}.

Since the beginning of the theory of $c \bar c$ system
\cite{charm1,charm2,charm3,charm4}, through the development
of potential models
\cite{pot1,pot2,pot3,pot4,pot5,pot6,pot7,pot8,pot9,pot10},
sum rules \cite{sumrules1,sumrules2}, and studies based on
effective Lagrangians \cite{nrqcd1,nrqcd2,nrqcd3}, the
latter method known also to work well in Hamiltonian
approach to QED \cite{krp1, krp2}, current understanding of
heavy quarkonia, especially in lattice approach
\cite{Wilsonlattice,nrLQCD,lattice2004}, is one of the best
examples of great progress achieved in theory of strong
interactions \cite{hqp}. In the wake of the development,
conceptual problems with a relativistic description of
hadrons in the Minkowski space and questions concerning the
structure of the vacuum state continue to bother theorists
\cite{Wilson2004,LFQCD}. The problem is how to derive a
quantum Hamiltonian for quarks and gluons that precisely
describes hadronic states in agreement with special
relativity. 

One can write a formal expression for a Hamiltonian of
quarks and gluons using various forms of dynamics
\cite{Dirac1}. In the standard form, one describes the
evolution of a system of particles using the time parameter
$t$, which is measured along a time-like direction in
space-time. But the formal expression must be regulated and
renormalized, and one has to explain how to define the
ground state of the theory that supports the tower of
excitations that represent all kinds of relativistically
moving and interacting hadrons. None of these tasks has been
completed yet with the desired clarity and precision. In the
LF form of dynamics, one uses a ``time'' parameter that is
conventionally denoted by $x^+ = t + z$. Since the LF
hyperplane $x^+ = 0$ is preserved by boosts along $z$-axis,
and also by two other boost-like transformations, the LF
form of Hamiltonian dynamics is invariant under the three
special boosts. The three boosts are prerequisite to the
construction of hadronic states in all kinds of motion. The
boost symmetry rises hope that the LF Hamiltonian dynamics
can help in finding a universal theoretical description of
hadrons in all frames of reference, including both their
center-of-mass frame (CMF), in which the constituent quark
model is developed \cite{PDG}, and the infinite momentum
frame (IMF), in which the concept of partons is developed
\cite{partons}. The exact description of a hadron in motion
is also essential in exclusive (or semi-exclusive)
processes~\cite{LB}. Regarding the vacuum, the ground state
problem in QCD does not appear in the LF form of dynamics in
the way known from the standard approach
\cite{frontreview1,LFQCD} and the LF Hamiltonian dynamics of
quarks and gluons is of great interest to many researchers
\cite{frontreview2}, as a serious alternative to the
standard form. However, LFQCD challenges theoreticians with
basic questions concerning quantum mechanics of particles
and fields and relativity.

In search for understanding of the quark-gluon dynamics, it
is natural to consider quantum chromodynamics of only heavy
quarks coupled to gluons. A quark is considered heavy when
the phenomenological mass parameter associated with the
quark, $m$, is much larger than $\Lambda_{QCD}$, the latter
being defined in the RGPEP procedure that one can use to
evaluate effective LF Hamiltonians \cite{RGPEP}. The
restriction to only heavy quarks creates a situation in
which the renormalization group parameter, denoted by
$\lambda$, can be simultaneously much smaller than $m$ and
much larger than $\Lambda_{QCD}$. In such circumstance, the
effective color coupling constant at scale $\lambda$,
$g_\lambda$, can be formally considered small in comparison
to 1 even when $\lambda$ is much smaller than $m$. The
smallness of the coupling constant implies that the LF
Hamiltonian of QCD expressed in terms of the effective quark
degrees of freedom corresponding to $\lambda \ll m$, denoted
by $H_\lambda$, can be evaluated in RGPEP using perturbation
theory. But the price to pay for this simplification is high
because the quenched heavy-flavor version of the theory must
be incomplete; it misses dynamical interplay among different
flavors and entirely ignores effects due to light quarks. On
the other hand, the price is worth paying because the single
heavy flavor theory quickly renders a simple dynamical
picture that may be helpful in learning more about LF QCD.
Namely, it is sufficient to consider $H_\lambda$ obtained in
just 2nd-order perturbation theory and augment it with an
ansatz for the mass gap for effective transverse gluons and
these two steps already render a boost-invariant eigenvalue
equation in the effective quark-antiquark Fock sector which
has a well-defined and phenomenologically attractive
structure \cite{ho}: the Coulomb potential with Breit-Fermi
corrections is supplemented with a harmonic oscillator term
with a frequency $\omega$ that is explicitly related to the
values of $\alpha$ and $m$ at the scales $\lambda$ at which
this structure may be valid, which turns out to be the scale
corresponding to the distances between quarks not much
larger than the size of the lowest-mass mesons. At larger
distances, additional gluons may be created and the
potential may become linear, as will be discussed later. 

The resulting dynamical picture for the low-mass states is
not sensitive to the details of the mass gap ansatz for
effective gluons that was used to finesse the picture.
Therefore, one needs to complete the 4th-order calculation
of $H_\lambda$ using RGPEP in order to begin a study of the
true mass gap for gluons that may undergird the finessed
effective picture. A major problem in the 4th-order
calculation is to include light quarks. Even before the
inclusion of light quarks, the first glimpse of the
magnitude of 4th-order effects could be obtained in
one-flavor QCD by comparison of the 2nd-order picture with
the 4th-order one. However, the 4th-order calculation
requires detailed understanding of many complex terms in the
LF Hamiltonian at once even in one-flavor QCD. It thus
becomes desirable to verify if the already known 2nd-order
picture can provide a reasonably accurate description of 
data and become a candidate for the first approximation that 
can organize the 4th-order studies.

Fortunately, in the limit of $m$ large in comparison to
$\lambda$ that itself is still much larger than
$\Lambda_{QCD}$, one can considerably simplify the
calculation of quark Hamiltonians using RGPEP. One can
compute a 2nd-order Hamiltonian that acts only in the
effective heavy quark-antiquark sector, $H_{Q \bar Q
\lambda}$, and evaluate interactions in $H_{Q \bar Q
\lambda}$ that depend on the quark spin. The result is that
the leading spin-dependent terms that one obtains from the
2nd-order RGPEP, in addition to the Coulomb potential and
the harmonic oscillator force, automatically respect
rotational symmetry. This leading simplest version of the
Hamiltonian approach can be used to calculate the spectrum
of masses of $c \bar c$ or $b \bar b$ mesons. Comparison
with experimental data is surprisingly optimistic. Not only
the masses that one hopes to be reproducible relatively well
turn out to be reproducible very well, but also those masses
that one expects to be reproducible rather poorly turn out
to be reproducible with considerably less accuracy than
those that are reproduced well. Theoretical results seem to
match data in a pattern that agrees with expectations
concerning accuracy of the simplest version of the approach.

Section \ref{biH} briefly describes the derivation of
the effective Hamiltonians studied here. The leading
approximation to the heavy quark eigenvalue problem is
explained in Section \ref{la}. Section \ref{spin}
discusses spin effects and lists eigenvalue equations
that describe mesons with spin 0, 1, and 2, with
orbital angular momenta equal 0, 1, 2, or 3. Masses and
wave functions of mesons that are obtained by numerical
solution of these equations are discussed in Section
\ref{masses}. Conclusions are summarized in Section
\ref{c}. Key details of the calculations are relegated
to Appendices.
\section{ Boost-invariant Hamiltonians } 
\label{biH} 
This section explains how the boost-invariant
Hamiltonians for heavy quarkonia that are used
in the next sections to evaluate masses of $c
\bar c$ and $b \bar b$ mesons are derived in
QCD. One uses the gluon mass-gap ansatz in the 
intermediate step of evaluating $H_{Q \bar Q \lambda}$
from the 2nd-order $H_\lambda$ \cite{ho}. 
\subsection{ Canonical LF QCD } 
One begins from the standard
Lagrangian of color gauge theory with one
flavor of quarks of mass $m$,
\begin{equation} 
{\cal L} = \bar \psi(i\hspace{-4pt}\not\!\!D - m)\psi -
{1\over 4}F^{\mu\nu a}F_{\mu\nu}^a \, .
\end{equation} 
The corresponding generator of
evolution in $x^+$ in the gauge $A^+=0$ takes
the form 
\begin{eqnarray} 
H_{can} &=&
H_{\psi^2} + H_{A^2} + H_{A^3} + H_{A^4} +
H_{\psi A \psi} \nonumber \\ 
&+& H_{\psi A A
\psi} + H_{[\partial A A]^2} + H_{[\partial A
A](\psi\psi)} + H_{(\psi\psi)^2},
\end{eqnarray} 
where each of the terms is an
integral of a corresponding Hamiltonian
density $\cal H$ over the LF hyperplane, $H_i
= \int dx^- d^2 x^\perp {\cal H}_i$. Four
terms that explicitly enter in the derivation
of the approximate 2nd-order boost-invariant
effective theory for heavy quarkonia, are
\begin{eqnarray} 
{\cal H}_{\psi^2} 
&=&
{1\over 2} \bar \psi \gamma^+
{-\partial^{\perp \, 2} + m^2 \over
i\partial^+} \psi \, , \\ {\cal H}_{A^2} &=&
- {1\over 2} A^\perp (\partial^\perp)^2
A^\perp \, , \\ {\cal H}_{\psi A \psi} &=& g
\, \bar \psi \hspace{-4pt}\not\!\!A \psi \, ,
\\ {\cal H}_{(\psi\psi)^2} &=& {1\over 2}g^2
\, \bar \psi \gamma^+ t^a \psi {1 \over
(i\partial^+)^2 } \bar \psi \gamma^+ t^a \psi
\, . 
\end{eqnarray} 
Other terms in the canonical Hamiltonian are 
also important. For example, the three-gluon 
coupling term plays an implicit role as a seed 
of the renormalization group flow of the coupling
constant. This effect becomes explicit first
in 3rd-order calculations~\cite{glambda}. 

At $x^+ = 0$, the fermion field 
\begin{eqnarray}
\label{Psi} 
\psi = \sum_{\sigma c} \int [k] \left[
\chi_c u_{k\sigma} b_{k\sigma c} e^{-ikx} + \chi_c
v_{k\sigma} d^\dagger_{k\sigma c} e^{ikx} \right] \, ,
\end{eqnarray} 
and the gluon field 
\begin{eqnarray}
\label{A} A^\mu = \sum_{\sigma c} \int [k] \left[ t^c
\varepsilon^\mu_{k\sigma} a_{k\sigma c} e^{-ikx} + t^c
\varepsilon^{\mu *}_{k\sigma} a^\dagger_{k\sigma c}
e^{ikx}\right] \, , 
\end{eqnarray} 
are quantized by
imposing commutation relations 
\begin{eqnarray}
\{b_{k\sigma c},b^\dagger_{k'\sigma' c'}\}
&=&\{d_{k\sigma c},d^\dagger_{k'\sigma' c'}\} \nonumber \\ 
\label{b} 
&=& 16\pi^3 k^+ \delta_{\sigma_\sigma'}
\delta_{cc'} \delta^3(k-k') \, , \\ 
\label{a} \left[
a_{k\sigma c}, a^\dagger_{k'\sigma' c'}\right] &=&
16\pi^3 k^+ \delta_{\sigma_\sigma'} \delta_{cc'}
\delta^3(k-k') \, . 
\end{eqnarray} 
The measure of integration over momenta, $[k]$, is
$\theta(k^+)$ $d k^+ \, d^2$ $k^\perp$ $/(16\pi^3 k^+)$ and
the LF three-momentum $\delta$-function is $\delta^3(k-k')$
= $\delta(k^+-k'^+)$ $\delta(k^1-k'^1)$ $\delta(k^2-k'^2)$.
Spins are denoted by $\sigma$ and colors by $c$. Further
details concerning our notation can be found in the
Appendix (see also Ref. \cite{ho}). 
\subsection{ Regularization } 

The canonical Hamiltonian is divergent and the RGPEP begins
with regularization of the ultra-violet and small-$k^+$
divergences. The ultraviolet divergences result from
integrating over large transverse momenta of quanta in the
intermediate states when one attempts to evaluate powers of
the Hamiltonian. In 3+1 dimensional theory, the
quadratically and logarithmically diverging transverse
integrals result from momentum dependent spin factors for
fermions and vector bosons. The small-$k^+$ divergences
arise due to gauge couplings of gluons. Note that in the
$A^+=0$ gauge only $A^\perp$ and $\psi^+ = {1 \over 2}
\gamma^0 \gamma^+ \psi$ are dynamical variables. In
particular, $A^-$ depends on $A^\perp$ and $\psi^+$. As a
consequence, interaction terms in the Hamiltonian can be
written using the polarization vector $\varepsilon^\mu$ for
gluons whose only two transverse components are independent
degrees of freedom. A gluon with momentum $k^+$ and
$k^\perp$ has $\varepsilon^- = 2 \, \varepsilon^\perp
k^\perp /k^+$. The $k^+$ in denominator in $\varepsilon^-$
is a source of small-$k^+$ singularities in LF QCD. There
exist similar small-$k^+$ singularities in the instantaneous
interactions along $z$-axis on the LF hyperplane, especially
where $1/\partial^{+ \, 2}$ appears, which happens similarly
to how the inverse of a three-dimensional Laplacian appears
in the familiar Coulomb potential in the standard approach.
The small-$k^+$ singularities also occur in a 1+1
dimensional theory \cite{tHooft}. 

In the boost-invariant formulation of LF QCD, one does not
regulate the theory by limiting the momentum components
$k^\perp$ and $k^+$ of every individual quark and gluon
separately. Instead, only the relative momenta of particles
in the interaction terms are limited. The regularization is
accomplished by insertion of regulating factors $r$ in
vertices, see \cite{ho}. The overarching rule for
construction of the regulating factors $r$ is that they must
respect 7 kinematical symmetries of the LF scheme. This rule
is analogous to the requirement of rotational symmetry in
the standard approach. There is a factor $r = r_\Delta
r_\delta$ for every bare particle in every vertex.
$r_\Delta$ limits the range of the relative transverse
momentum of an interacting bare particle with respect to
other particles participating in the interaction. The limit
is set by the parameter $\Delta$. If a particle of momentum
$k$ carries a fraction $x$ of the total $p^+$ of particles
in an interaction term, $k^+ = x p^+$, then it's relative
transverse momentum with respect to the particles
interacting in this term is defined as $\kappa^\perp =
k^\perp - x p^\perp$, where $p^\perp$ is the sum of
transverse momenta of all particles annihilated or
(alternatively) created in the term. One uses $r_\Delta =
\exp{(- \kappa^{\perp \, 2}/\Delta^2)}$, and the ultraviolet
regularization parameter $\Delta$ is sent to infinity in
comparison to all physical momentum scales, cf.
\cite{Thorn1,Thorn2,Thorn3,Thorn4}.

In the case of the small-$k^+$ singularities, the regulating
factors $r_\delta$ limit the ratio $x = k^+/p^+$ by the
positive arbitrarily small dimensionless parameter $\delta$.
All that is required of the factors $r_\delta$ is that they
vanish as $x^\delta$ when $x \rightarrow 0$. This condition
is sufficient in globally colorless states. Linear
divergences at small $x$ cancel out and one only needs to
take care of the logarithmic divergences in integrals of the
type $\int dx/x$. The small-$x$ divergences in the gauge
boson dynamics occur in both ultraviolet and infrared
regimes. Massless particles can simultaneously have small
$x$ and small $\kappa^\perp$ and their virtuality in the
small-$x$ region can be large or small depending on the
ratio of $|\kappa^\perp|$ to $\sqrt{x}$. Small $x$ implies
large virtuality only for particles with non-zero mass or
fixed $\kappa^\perp$. 

Once the canonical Hamiltonian is regulated, one needs to
introduce counterterms that restore the physics that existed
outside the cutoff range. Thus, the counterterms remove
effects of the regularization. For example, one inserts mass
and vertex counterterms and they remove dependence on the
artificial ultraviolet regularization factors $r_\Delta$.
The resulting regulated Hamiltonian with counterterms,
\begin{eqnarray}
\label{H}
H = \left[ H_{can} + H_{CT} \right]_{reg} \, , 
\end{eqnarray}
provides the starting point for further steps. The further
steps are also helpful in establishing the structure of the
required counterterms. Note that the dynamics of
color-singlet states of finite size should not be sensitive
to the small-$x$ regularization. Namely, the singular limit
$x \rightarrow 0$ concerns gluons with long wavelengths in
the direction of $x^-$. But the strength of the coupling of
such gluons to a finite-size color-neutral pair of quarks
should disappear when the wavelength becomes infinitely
larger than the distance between the quarks.
%
\subsection{ Effective particles }
%
The initial Hamiltonian $H$ of Eq. (\ref{H}) is expressed in
terms of the creation and annihilation operators defined by
the Fourier components of local fields in Eqs. (\ref{Psi})
and (\ref{A}). The same $H$ can be expressed in terms of
creation and annihilation operators for effective quarks and
gluons that correspond to a renormalization group scale
$\lambda$ in RGPEP. The procedure is constructed in such a
way that the operators return to the canonical operators
when $\lambda$ tends to infinity, but when $\lambda$ is
near the energy-scale of the binding mechanism, on the order
of masses of hadrons, the operators create or annihilate
effective particles that are expected to correspond to the
constituent quarks and gluons. The quantum numbers of the
constituents are the same as in the local theory and one
assumes that the corresponding creation and annihilation
operators are related by a unitary transformation
\begin{eqnarray}
q_\lambda & = & U_\lambda \, q_{can} \, U_\lambda^\dagger \, , 
\end{eqnarray}
where the same letter $q$ is used for both creation and
annihilation operators. The next step is to express $H$
in terms of $q_\lambda$ instead of $q_{can}$,
\begin{eqnarray}
H_\lambda (q_\lambda) & = & \left[ H_{can} + H_{CT} \right]_{reg}(q_{can}) \, .
\end{eqnarray}
The Hamiltonian $H$ remains the same but the
coefficients in the expansion in powers of $q_\lambda$
are new. They include potentials whose structure can be
calculated order-by-order in perturbation theory using
the RGPEP. The key feature is that $H_\lambda$ has the
structure \cite{fG}
\begin{equation}
\label{fG}
H_\lambda = f_\lambda G_\lambda \, ,
\end{equation}
where $f_\lambda$ denotes form factors of width
$\lambda$ and $G_\lambda$ represents interaction
vertices that can be calculated for any assumed shape
of $f_\lambda$. The shape we use here is best described
using the example of a term in which an effective quark
emits an effective gluon:
\begin{eqnarray}
G_\lambda 
& = & 
\int [123] \, G_\lambda(1,2,3) \, 
a^\dagger_{\lambda 1} b^\dagger_{\lambda 2} b_{\lambda 3}   \, , \\
f_\lambda G_\lambda 
& = & 
\int [123] \, f_\lambda(123) \, G_\lambda(1,2,3) \, 
a^\dagger_{\lambda 1} b^\dagger_{2 \lambda}  b_{\lambda 3} \, , \\
\label{fl}
f_\lambda(123) 
& = & 
\exp{\left[ - ({\cal M}_{12}^2 - {\cal M}_3^2)^2/\lambda^4 \right]} \, .
\end{eqnarray}
The invariant masses are defined by the formulas ${\cal
M}_{12}^2 = (k_1 + k_2)^2$, ${\cal M}_3^2 = k_3^2$,
using masses in the Lagrangian with $g=0$ to evaluate
minus components of the four-momenta; $m$ for quarks 
and 0 for gluons.

The operator $G_\lambda$ is defined by the coefficients
of its expansion in a series of powers of operators
$q_\lambda$. One can also define ${\cal G}_\lambda$,
which is a series with the same coefficients but
$q_\lambda$ replaced by $q_{can}$. Then, one can use
the constant operator basis $q_{can}$ when solving
differential equations of RGPEP for the coefficients,
see \cite{RGPEP}. ${\cal G}_\lambda$ is split into two
parts: ${\cal G}_0$ and ${\cal G}_I = {\cal G} - {\cal
G}_0$. ${\cal G}_0$ is the part that does not depend on
the coupling constant $g$. The RGPEP differential
equation for ${\cal G}_I$ (prime denotes
differentiation with respect to $\lambda$) is:
\begin{equation}
\label{G}
{\cal G}_I'\,\, = \,\, \left[ f{\cal G}_I, \, \left\{ [(1-f){\cal
G}_I]' \right\}_{{\cal G}_0} \right] \, ,
\end{equation}
where the curly bracket with subscript ${\cal G}_0$ is
introduced to indicate the operator ${\cal T}$ that
solves equation $[{\cal T}, \, {\cal G}_0] = [(1 - f)
{\cal G}_I]'$. The initial condition is that ${\cal
G}_\infty = H$, and the solution is
\begin{equation}
\label{intG}
{\cal G}_\lambda \,\, = \,\, H \,\, + \,\,
\int_\infty^\lambda ds \,\left[ f_s{\cal G}_{Is}, \, \left\{[(1-f_s){\cal
G}_{Is}]' \right\}_{{\cal G}_0} \right] \, .
\end{equation}
This solution is evaluated order-by-order in powers of the
coupling constant $g_\lambda$ that appears in the
vertices of $G_\lambda$~\cite{glambda}. The
perturbative expansion is legitimate because one never
encounters small energy denominators. This feature is
secured by the structure of the RGPEP equations and the
shape of the form factor $f$ (see the original
literature). $G_\lambda$ is obtained from ${\cal
G}_\lambda$ by replacing $q_{can}$ by $q_\lambda$. 

Solving Eq. (\ref{intG}) up to terms of order $g^2$ (there
is no difference between the expansions in powers of the bare
coupling constant $g$ and the running coupling constant
$g_\lambda$ in 2nd-order terms, but one should think about
the expansion in powers of $g_\lambda$, see below and next
sections), which includes finding the mass counterterms in
the initial condition at $\lambda = \infty$, one obtains
$H_\lambda$ that can change the number of effective particles 
by 0, 1, or 2. For 2nd-order evaluation of the effective 
Hamiltonian in the quark-antiquark sector, $H_{Q \bar Q 
\lambda}$, one only needs the following terms \cite{ho}
\begin{equation}
\label{Hl}
H_\lambda = T_{q \lambda} + T_{g \lambda} 
          + f_\lambda \left[ Y_{qg \lambda} 
          + V_{q \bar q \lambda} + Z_{q\bar q \lambda} \right] \, .
\end{equation} 
$T_{q \lambda}$ and $T_{g \lambda}$ denote the kinetic
energy operators for quarks and gluons, respectively.
$Y_{qg \lambda}$ denotes the term which causes that
effective quarks emit or absorb effective gluons (the
letter $Y$ is chosen in the notation because its shape
resembles an act of one particle splitting into two, or
two particles forming one). $f_\lambda V_{q \bar q
\lambda}$ is an interaction between quarks due to
exchange of gluons with virtuality greater than
$\lambda$. $f_\lambda Z_{q\bar q \lambda}$ is the
instantaneous interaction between effective quarks that
originates in the instantaneous interaction in the
canonical LF Hamiltonian. Details of these terms are
listed in Appendix \ref{TermsinHlambda}.
%
\subsection{ Derivation of $H_{Q \bar Q \lambda}$ }
%
This section explains how one obtains the effective
Hamiltonian $H_{Q \bar Q \lambda}$ for a heavy
quarkonium starting from the eigenvalue problem for the
effective Hamiltonian $H_\lambda$ that reads
\begin{equation}
\label{evh}
H_\lambda |P\rangle \, = \,  E |P\rangle \, .
\end{equation}
$|P\rangle$ denotes an eigenstate of the operators
$P_\lambda^+$ and $P^\perp_\lambda$ with eigenvalues
$P^+$ and $P^\perp$ (see \cite{algebra} for an example
of RGPEP construction of the whole Poincar\'e algebra).
The eigenvalue has the form $E \, = \, (M^2 + P^{\perp
\, 2}) /P^+$ and one obtains an eigenvalue equation for 
$M^2$ by multiplying Eq. (\ref{evh}) by $P^+$ and 
subtracting $P^{\perp \, 2}$. $|P\rangle$ is expanded 
in the effective particle basis as
\begin{equation}
\label{|P>}
|P\rangle = |Q_\lambda \bar Q_\lambda \rangle + 
            |Q_\lambda \bar Q_\lambda g_\lambda \rangle + 
            \, . \, . \, . \, \, .
\end{equation}
For $\lambda$ much smaller than $m$ this expansion is
dominated by its components with only two heavy quarks
because the vertex form factors $f_\lambda$ in
$H_\lambda$ eliminate the probability of creating
components with invariant masses that differ from $2m$
by much more than $\lambda$. One may also expect that
gluons develop a mass gap in QCD and the components
with many gluons are also suppressed. If one neglects
sectors with effective gluons entirely, the eigenvalue
problem is reduced to
\begin{eqnarray}
\label{evqq}
\left[ T_{q \lambda} + f_\lambda 
\left( V_{q \bar q \lambda} + Z_{q\bar q \lambda}\right) 
\right]|Q_\lambda \bar Q_\lambda \rangle = 
E |Q_\lambda \bar Q_\lambda \rangle \, .
\end{eqnarray}
But Eq. (\ref{evh}) implies that the $|Q_\lambda \bar
Q_\lambda g_\lambda \rangle$ component satisfies equation
\begin{eqnarray}
\label{evqqg}
\left[ T_{q \lambda} + T_{g \lambda} + 
       V_{q \bar q g \lambda} - E \right]  
      |Q_\lambda \bar Q_\lambda g_\lambda \rangle =
- Y_\lambda  |Q_\lambda \bar Q_\lambda \rangle \, , 
\end{eqnarray}
and can contribute to the dynamics in the sector $ |Q_\lambda
\bar Q_\lambda \rangle $ in order $g_\lambda^2$, or
$\alpha_\lambda = g_\lambda^2/(4\pi)$, because $Y_\lambda$
is of order $g_\lambda$. $V_{q \bar q g \lambda}$ denotes
potentials in the three-body sector, including non-Abelian
potentials that act between the effective gluon and quarks.
Additional interactions with sectors that contain four or
more effective particles are not indicated. The additional
interactions and $V_{q \bar q g \lambda}$ are expected to
cause a shift in the gluon energy and make the eigenvalue
equation differ from a similar one for positronium. In
positronium, a state with two or more photons could have the
same energy as the state with one photon. In QCD, there
exist potential terms that act between gluons and quarks and
among gluons themselves that have no counterpart in QED.
It is very unlikely that there does not exist some shift in
gluon energy that is absent in the case of photons. One can
employ an ansatz for the effective gluon mass in the
three-body sector to study possible consequences of such
shift \cite{ho}.

The point is that one can study the dynamics of $H_\lambda$
order by order in $g_\lambda$ using a scheme of successive
approximations that include an ansatz for effects that are
extremely small for an infinitesimal $g_\lambda$ but need to
be included to come close to a true solution that is
obtained only when the coupling constant takes values
comparable with 1. In each successive order one can replace
the ansatz terms introduced in a lower order by a true
interaction of that lower order but with the coupling
constant in them extrapolated to the large physical
value~\cite{LFQCD}. The task of finding the initial ansatz
terms that come close to the actual dynamics with large
relativistic coupling constant may in principle require a
lot of research to complete. Fortunately, the
boost-invariant effective particle approach has a useful
feature: a lowest-order ansatz that is defined using an
effective mass-like term for constituent gluons allows one
to easily calculate and eliminate the gluon component and
the resulting dynamics of constituent quarks comes out
rotationally symmetric and independent of the details of the
gluon mass ansatz one makes provided only that the ansatz
satisfies some general conditions \cite{ho}. In this first
approximation, all interactions in the sector $|Q_\lambda
\bar Q_\lambda g_\lambda \rangle$ are replaced by $\mu^2$,
which is a function of the relative motion of the three
constituents. $\mu^2$ must vanish when the gluon $x
\rightarrow 0$. Since the mass ansatz for $\mu^2$ is
supposed to model the dominant effect of all the
interactions within the three-body sector and with sectors
of larger numbers of constituent particles when the coupling
constant takes a realistically large value, the first term
in the ansatz can be considered to be on the order of 1 in
comparison to the terms that depend on the infinitesimal
coupling constant used in the formal expansion in RGPEP. The
whole eigenvalue problem for $H_\lambda$ with the ansatz is
now reduced to only two coupled equations ($\lambda$ is
omitted), 
\begin{eqnarray} 
\label{matrix}
(T_q + \tilde T_g) |Q\bar Qg\rangle + Y|Q\bar Q\rangle
& = & E |Q\bar Qg\rangle , \\ Y |Q\bar Qg\rangle +
\left[T_q + f\left( V_{q \bar q} + Z_{q\bar
q}\right)\right] |Q\bar Q\rangle & = & E |Q\bar
Q\rangle . 
\end{eqnarray} 
The operator $\tilde T_g$ is marked with the tilde in order
to indicate that the effective gluon mass $\mu_\lambda^2$ 
in Eq. (\ref{tgl}) is replaced by the ansatz mass $\mu^2$ 
in the 3-body sector.

The Hamiltonian $H_{Q\bar Q}$ that acts only in the $|Q
\bar Q \rangle$ sector can now be evaluated as a power
series in $g$ using an operator usually denoted by $R$
\cite{R}. In the simplest version, $R$ expresses the 3-body
component through the 2-body one, $|Q\bar Qg\rangle =
R|Q\bar Q\rangle$. Note that the sector $|Q \bar Q g \rangle$ 
is separated from the sector $|Q \bar Q \rangle$ by a gap in 
invariant mass. The 2nd-order result is the Hamiltonian 
whose matrix elements are \cite{ho}
\begin{eqnarray}
\label{hqqij}
&& \langle 13|H_{Q\bar Q}|24\rangle = \langle 13|
\left[ T_q\, + f\left( V_{q \bar q} + Z_{q\bar q}\right) \right] 
|24\rangle + \nonumber \\
&& \langle 13| fY_{qg}\left[ {1/2 \over E_{24}-T_q -\tilde T_g} +
                      {1/2 \over E_{13}-T_q - \tilde
T_g} \right] fY_{qg} |24\rangle \, ,
\nonumber 
\end{eqnarray}
where $|ij\rangle$ with $i$ equal 1 or 2 and $j$ equal
3 or 4 are eigenstates of the operator $T_q$ in the $|Q
\bar Q \rangle $ sector of the Fock space, and $E_{ij}$
are the corresponding eigenvalues. The labeling of
states is illustrated in Fig. \ref{fig:oge}. The basis
states are defined as \begin{eqnarray} |ij\rangle & = &
b^\dagger_{\lambda i} d^\dagger_{\lambda j} \,
|0\rangle \, , \end{eqnarray} where $b^\dagger_\lambda$
and $d^\dagger_\lambda$ are creation operators for
effective quarks and antiquarks corresponding to the
RGPEP width parameter $\lambda$. The corresponding
eigenvalue is $E_{ij} = ({\cal M}_{ij}^2 + P^{\perp \,
2})/P^ +$, where ${\cal M}_{ij}^2 = (k_i + k_j)^2$ and
the minus components of the four-momenta are evaluated
as for free particles of mass $m$.

In summary, the procedure used here \cite{ho} 
replaces the eigenvalue problem for $H_\lambda$
by an eigenvalue problem with an ansatz (dots 
denote operators that couple states with more 
effective particles than three)
\begin{eqnarray}
\label{ansatzscheme1}
\left[H_\lambda\right] & = & 
\left[
\begin{array}{ccc}
\cdot~      &  \cdot      &  \cdot     \\
\cdot~      &  H_3        &  Y         \\
\cdot~      &  Y^\dagger  &  H_2 
\end{array}
\right]
\rightarrow 
\left[
\begin{array}{cc}
T_3 + \mu^2 &  Y  \\
Y^\dagger   &  T_2 + V_2
\end{array}
\right] \, ,
\end{eqnarray}
and then the operation $R$ is used to derive the
effective quark Hamiltonian (in a simplified 
symbolic notation) 
\begin{eqnarray}
\label{ansatzscheme2}
H_{Q\bar Q \lambda } & = & 
T_{2\lambda} + V_{2\lambda} + 
Y_\lambda^\dagger  {1 \over T_3 + \mu^2}  Y_\lambda \, .
\end{eqnarray}
The procedure should not be confused with a conventional
Tamm-Dancoff approach to quantum field theory. The quantum
particle degrees of freedom that are obtained from RGPEP are
not the bare quanta of local canonical theory, see
\cite{largep}, and the effective particles obey rules of the
LF dynamics with a vacuum that is simple to work with.
Moreover, the effective particles interact through terms
like $Y_\lambda$ that contain vertex form factors whose
width has interpretation of the size of the effective
particles in strong interactions (the particles cannot emit
or absorb any quanta with greater invariant mass changes
than $\lambda$). At the same time, $\lambda$ also plays the
role of the RGPEP parameter in the differential equations
that control the evolution of operators from the canonical
ones at $\lambda=\infty$ to the effective ones that can be
used in a relativistic computation of bound states when
$\lambda$ is lowered to the scale of the hadronic masses.
The relativistic nature of the procedure is reflected by the
possibility to construct all generators of the Poincar\'e
group at the scale $\lambda$ at which one wishes to solve
the eigenvalue problem for $H_\lambda$ itself
\cite{algebra}. 

Although the eigenvalue problem for heavy quarkonia in QCD
with explicit inclusion of the sector with 3 effective
particles has not been studied in detail yet, it is
important to state here that the ansatz scheme dictates in
this case the replacement 
\begin{eqnarray}
\label{ansatzscheme3}
\left[
\begin{array}{cccc}
\cdot~      &  \cdot       &  \cdot        &  \cdot   \\
\cdot~      &  H_4         &  Y_1          &  Y_2     \\
\cdot~      &  Y_1^\dagger &  H_3          &  Y       \\
\cdot~      &  Y_2^\dagger &  Y^\dagger    &  H_2
\end{array}
\right]
\rightarrow 
\left[
\begin{array}{ccc}
T_4 + \mu^2 &  Y_1          &  Y_2     \\
Y_1^\dagger &  H_3          &  Y       \\
Y_2^\dagger &  Y^\dagger    &  H_2
\end{array}
\right]
\end{eqnarray}
and subsequent application of $R$ to the desired order.
These steps appear to resemble the LF Tamm-Dancoff scheme
with sector-dependent counterterms proposed by Perry,
Harindranath, and Wilson \cite{PHW}. The conceptual
difference is that in the Perry-Harindranath-Wilson scheme
the elimination of sectors occurs within a Hamiltonian
eigenvalue problem with large cutoffs and the ultraviolet
renormalization issue is a part of the problem, leading to
Wilson's triangle of renormalization with a vast space of 
relativistic quantum operators. In the scheme used here 
\cite{ho}, the ultraviolet renormalization group procedure 
is completed long before one tackles the eigenvalue problem 
and introduces an ansatz at the scale $\lambda$ near the 
magnitude of invariant masses that characterize observables. 
The principles of extracting a small and computer-soluble 
eigenvalue problem from an eigenvalue problem of infinite 
size, such that the results obtained from a small problem 
can represent solutions to the infinite problem, will be 
further explained below.

Finally, it should be stressed that the mass ansatz
has the structure \cite{ho}
\begin{eqnarray}
\mu^2_{ansatz} & = & \left[ 1 - \alpha_\lambda^2 / \alpha_s^2 \right] \, \mu^2 \, , 
\end{eqnarray}
where $\alpha_s$ denotes the large, relativistic coupling
constant that the effective coupling $\alpha_\lambda$ is
supposed to reach as a result of extrapolation from the
infinitesimal values used in the perturbative solution of
the equations of RGPEP for $H_\lambda$. This structure 
ensures that in the lowest-order approximate expressions 
obtained through expansion in powers of $\alpha_\lambda$, 
only the first term counts,
\begin{eqnarray}
\mu^2_{ansatz} & = & \mu^2 \, . 
\end{eqnarray} 
This term is order 1. But when one increases the order of
included terms and extrapolates to $\alpha_\lambda =
\alpha_s$, the ansatz is removed and one has a chance to
recover a true solution of the initial theory with
increasing accuracy. Comparisons of results obtained from
expansions in successive orders, and use of better
extrapolation techniques than through a plain power series,
will hopefully indicate if the procedure can converge on a
well-defined dynamical picture. For that purpose, one should
compare the initial approximate theories with experiment and
find out if the coupling constants required for explaining data
can be small enough to pursue the chain of calculations
based on the weak-coupling expansion. Results of this study
suggest that at least for heavy quarkonia the required
coupling constant in LF QCD is considerably smaller than 1,
see below.
%
\subsection{ Eigenvalue equation for $H_{Q \bar Q \lambda}$ }
%
\label{states}
The eigenvalue equation for $H_{Q \bar Q \lambda}$ has
the form
\begin{eqnarray}
\label{evhQQ}
H_{Q \bar Q \lambda} |Q_\lambda \bar Q_\lambda \rangle 
& =  & 
{ M^2 + P^{\perp \, 2} \over P^+} |Q_\lambda \bar Q_\lambda \rangle \, .
\end{eqnarray}
The eigenstates are written as (see Appendices
\ref{TermsinHlambda} and \ref{RGPEPscaling} for details
of the notation, subscript $\lambda$ is omitted)
\begin{eqnarray}
\label{state}
|M,P^+, P^\perp \rangle 
& = & \int [ij] \, \tilde \delta \, P^+  \, 
{\bar u_i \Psi_{ij} v_j \over - 4m^2} \, |ij\rangle \, .
\end{eqnarray}
The eigenstate wave function can be written in the form
that exhibits its covariance under 7 kinematical LF
Poincar\'e transformations, 
\begin{eqnarray}
\label{wavefunction}
\Psi_{ij} & = & 
\sum_{s_i s_j}  \Psi_{s_i s_j}(\vec k_{ij}) \,  u_{k_i, s_i} \, \bar v_{k_j, s_j} \, , \\
\label{wavefunctionCMF}
\Psi_{s_i s_j}(\vec k_{ij}) & = & 
\bar u_{\vec k_{ij}, s_i} \, \Psi_{CMFij}(\vec k_{ij}) \, v_{-\vec k_{ij}, s_j} \, ,
\end{eqnarray}
where $\Psi_{CMFij}(\vec k_{ij})$ denotes the wave
function that depends on the relative three-momentum of
quarks in their CMF, assuming their masses are just
$m$. The indices $s_i$ and $s_j$ denote projections of
spin on $z$-axis. Spinors $u_{k_i, s_i}$ and $v_{k_j,
s_j}$ are obtained using LF boost matrix ($\Lambda^\pm
= \gamma^0 \gamma^\pm/2$)
\begin{eqnarray}
\label{Bkm}
B(k,m) & = & 
{1 \over \sqrt{k^+ m}} \left[ k^+ \Lambda^+ + \Lambda^- ( m + k^\perp \alpha^\perp ) \right] 
\end{eqnarray}
acting on the spinors at rest, $u_{0,s}$ and $v_{0,s}$, in
the reference frame in which the bound state calculation is
carried out and where the four-momentum of the bound state
has components $P^+$, $P^\perp$, and $P^- = (M^2 + p^{\perp
\, 2})/P^+$, $M$ being the eigenvalue that one wants to
calculate. Spinors $u_{\vec k_{ij}, s_i}$ and $v_{-\vec
k_{ij}, s_j}$ are obtained by ``boosting'' spinors at rest
in the CMF of the constituent fermions along their relative
three-momentum. An additional spatial rotation is applied to
spinors in the CMF before the latter boost is applied, in
order to build a spin basis in which one obtains explicit
rotational symmetry of spin-dependent interactions in the
leading approximation. The additional rotation is the same
as the well-known Melosh transformation \cite{Melosh1,
Melosh2}. Details of our notation for momentum variables,
spinors, and boost matrices, are explained in Appendices
\ref{RGPEPscaling} and \ref{additionalturn}. 

Eq. (\ref{evhQQ}) implies the eigenvalue equation for
the wave function,
\begin{eqnarray}
0 & = & 
\left[
{\kappa_{13}^{\perp\,2} + m_\lambda^2 \over x_1 x_3 }
+
{m_{Y1}^2 \over x_1}
+
{m_{Y3}^2 \over x_3}
-
M^2
\right] \,
\Psi_{s_1 s_3}(\vec k_{13}) \nonumber \\
& - &
{{4 \over 3} g^2 \over 16\pi^3}
\int 
{dx_2 d^2 \kappa_{24}^\perp \over x_2 x_4 } \sum_{s_2 s_4}
v_\lambda (13,24) \, \Psi_{s_2 s_4}(\vec k_{24}) \, ,
\end{eqnarray}
in which the mass-like terms $m_{Y1}^2$ and $m_{Y3}^2$
result from the self-interaction of effective quarks through
emission and re-absorption of effective gluons, and
$v_\lambda(13,24)$ results from the exchange of the
effective gluon between the two quarks. The
ultraviolet-finite part of the mass counterterm in the
effective quark mass $m_\lambda^2$ is so adjusted (using a
single quark eigenvalue problem) that at $\lambda =
\lambda_0$ one obtains \cite{ho}
\begin{eqnarray}
\label{MainEq}
& & 
\left[
{\kappa_{13}^{\perp\,2} + m^2 \over x_1 x_3 }
+
{\delta m^2_1 \over x_1}
+
{\delta m^2_3 \over x_3}
-
M^2
\right] \,
\Psi_{s_1 s_3}(\vec k_{13})   \nonumber \\
& &
-
{{4 \over 3} g^2 \over 16\pi^3}
\int 
{dx_2 d^2 \kappa_{24}^\perp \over x_2 x_4 } \sum_{s_2 s_4}
v_0 (13,24) \, \Psi_{s_2 s_4}(\vec k_{24}) = 0 \, , \nonumber \\
& &  
\end{eqnarray}
where 
\begin{eqnarray}
\label{v0}
v_0(13,24) & = & - A \,g_{\mu \nu} \,
j_{12}^\mu \bar j_{43}^\nu 
+
B \, { j_{12}^+ \bar j_{43}^+ \over P^{+ \, 2} } \, , 
\end{eqnarray}
and $\delta m_1^2$, $\delta m_3^2$, $A$, $B$, and other
symbols, are explained in detail in Appendix \ref{RGPEPscaling}.
%
\section{ Leading approximation }
\label{la}
The RGPEP result of Eq. (\ref{MainEq}) is further analyzed
as a typical window Hamiltonian eigenvalue problem of the
kind studied in detail in the case of a generic matrix model
with asymptotic freedom and a bound state
\cite{modelafbs1,modelafbs2,modelafbs3}. The model is
soluble exactly and provides a relatively well-understood
pattern to follow in the case of QCD with one heavy flavor.
Earlier LF studies, based on coupling coherence
\cite{gcoh1,gcoh2,gcoh3,QQgcoh,QQ}, did not have such a
pattern to follow and did not use a boost-invariant concept
of effective particles. They were carried out in a frame of
reference nearly at rest with respect to the CMF of the
quarkonium, and employed a logarithmically confining
potential that was obtained in the quark-antiquark sector
neglecting all other sectors. Those studies pioneered an
attack on the bound-state problem in LF QCD along the path
discussed by Perry \cite{gcoh2,gcoh3}, including elements of
the method outlined in Wilson et al. \cite{LFQCD}, such as
the absolute cutoffs on momentum variables (especially
$k^+$) in the Fourier analysis of field variables in
position space, or similarity RG procedure for Hamiltonians.
One of the key issues of the LF approach is how to obtain
rotational symmetry and the initial studies had to struggle
with the issue, in addition to the issue of construction of
counterterms that restore boost symmetry violated by cutoffs
on absolute momentum variables. The RGPEP procedure used
here leads in its simplest version to a boost-invariant and
rotationally symmetric spectrum of meson masses.
%
\subsection{ Coupling constant in the window }
%
Construction of a window eigenvalue problem begins with
a selection of a set of states of effective particles
with kinetic energies (actually, free invariant masses) 
in a certain range that is also called a window, for 
brevity. The size of this range should be larger than 
the width $\lambda_0$ which appears in the form factors 
$f$ in $H_{\lambda_0}$. 

The next step is to evaluate matrix elements of
$H_{\lambda_0}$ in the selected window of basis states.
These matrix elements form a matrix $W$ of the window
Hamiltonian whose non-perturbative diagonalization is
to produce the bound state of interest. To facilitate 
efficient diagonalization in a continuum theory, one can 
use a set of orthonormal wave packets (such as the wave 
functions that solve a two-body bound-state eigenvalue 
problem with a harmonic oscillator potential) as a basis 
in which the matrix elements of the window are evaluated.

Typically, if the energy range (the word ``energy'' should
be replaced by the words ``invariant mass'' almost
everywhere in this paper, but the reader is expected to be
more familiar with the word energy than invariant mass in
reference to the quantum dynamical concepts that count here
and we use the word energy in order to avoid confusion due
to the lack of familiarity with the LF form of quantum
Hamiltonian dynamics) in the window Hamiltonian matrix $W$
is sufficiently larger than $\lambda_0$, the middle
eigenvalues of the window are independent of the window
boundaries and they match the eigenvalues of the full
$H_{\lambda_0}$. The latter eigenvalues are equal to the
exact eigenvalues of the initial Hamiltonian $H$ if the
RGPEP procedure is carried out exactly. The additional
virtue of lessons from Ref. \cite{modelafbs3}, beyond
showing that an asymptotically free model can be solved
using a window, is that one can also evaluate the matrix
elements of $W$ in perturbation theory, as if the coupling
constant was extremely small. One sets the coupling constant
to a realistically large value when one solves the
non-perturbative eigenvalue problem for $W$.

The point is that a few low orders in RGPEP calculation of
$W$ may lead to a good approximation (reaching better
accuracy than 10\% for the bound-state eigenvalues already
when $W$ is calculated in second order) if one properly
chooses $\lambda_0$ in order to work with a small number of
basis states (small means small enough so that they can be
handled using computers) and if one adjusts the coupling
constant in the window to the chosen $\lambda_0$. $\lambda_0$
should be near the scale of invariant masses that dominate
in the binding mechanism. The coupling constant is
defined through the value of a specific matrix element in
the window. It is adjusted by comparison of the spectrum of
$W$ with data (in \cite{modelafbs3}, the role of data is
played by the known exact spectrum). The main result of the
matrix model (studied so far up to 6th order, or 5 loop
integrals) is that when the coupling constant in the window
$W$ is adjusted so that one middle eigenvalue of $W$ matches
the corresponding exact solution then also other middle
eigenvalues of $W$ approximate the corresponding exact
solutions. 

In a theory as complex as QCD, the RGPEP calculations of
window Hamiltonians beyond 2nd order will require a lot of
work. Completion of the 4th order calculation is important
for determination of the accuracy one can achieve using
window Hamiltonians in QCD. Apart from the plain
perturbative expansion, one may eventually take advantage
of the idea of coupling coherence \cite{gcoh1}, reformulated
for the use in RGPEP. However, when additional flavors of
quarks are included and their masses are lowered toward
small values known in the standard model, one may have to
deal with an infrared limit cycle \cite{irlcQCD} (and
universality that may apply in that case
\cite{universality}, instead of the asymptotic freedom
structure known in the ultraviolet). But in the case of
heavy quarks, i.e., when the quark mass is formally very
large in comparison to $\Lambda_{QCD}$, the value of the
coupling constant required in the window may be small and no
complications possible for light quarks are expected to
occur.
%
\subsection{ Heavy quark limit }
%
In a formal analysis of QCD with one heavy flavor, the quark
mass $m$ can be much larger than $\lambda_0$ and the latter
much larger than $\Lambda_{QCD}$. In these circumstances,
the perturbative coupling constant corresponding to
$\lambda_0$ is small and the relative motion of quarks in
the sector $| Q \bar Q \rangle$ is limited by the form
factors $f$ of width $\lambda_0$. This means that the
dominant relative momenta of quarks in mesons are small in
comparison to $m$. (One should observe that the process of
extrapolation of the window to a large value of the coupling
constant corresponds to the increase in the value of
$\Lambda_{QCD}$; at a phenomenologically useful value of
$\alpha_0= \alpha_{\lambda_0}$, one may have to work with
$\lambda_0$ that is no longer very small in comparison to
$m$). Formally, in the limit of large $m$, the dominant
dynamical effects in the window eigenvalue Eq.
(\ref{MainEq}) with $\lambda_0 \ll m$ can be analyzed using
a non-relativistic approximation for the relative motion of
quarks. The arbitrary motion of the meson as a whole can be
exactly separated from the relative motion of the
constituent quarks because the RGPEP is invariant under the
LF boosts. Therefore, approaching the IMF, one will control 
the differences between absolute values of momenta of the
quantum constituents that extend to much larger values than
the mass of the entire meson. The differences correspond to
some fixed range of Feynman $x$ around 1/2. But every fixed
value of $x$ different from 1/2 implies that the corresponding
quark momenta in the IMF differ by amounts infinitely larger
than any fixed mass parameter.

The non-relativistic approximation is formally validated by
the condition that $\lambda_0 \ll m$, and that the
interaction terms are growing not faster than a polynomial
of kinetic energy (invariant mass) and cannot overcome the
exponential falloff of the form factors $f$ for changes of
invariant masses that are larger than $\lambda_0$. Thus, in
order to use the non-relativistic expansion, one has to keep
in place the exponential form factors $f$ that provide the
convergence - these form factors determine the size of the
window in momentum space and one cannot expand them in a
series. What can be expanded are the arguments of the
exponential functions, the perturbative factors that appear
in RGPEP in addition to the form factors, spin dependent
factors in interaction vertices, and relativistic measure of
integration over relative momenta of quarks. The accuracy of
results of diagonalization of the window $W_{\lambda_0}$
will depend on the choice for the form factor function $f$
and, especially, on the optimization factors that are
critical for the convergence of the perturbative evaluation
of $H_{\lambda_0}$ in 4th order
\cite{modelafbs2,modelafbs3}. The optimization factors were
considered in the case of heavy quarkonia elsewhere
\cite{JMlynikPhD}. The study described here was carried out
using $f$ of the generic type exemplified in Eq. (\ref{fl}).
Every form factor considered here is the same exponential
function of the square of a difference of squares of
invariant masses of the effective particles in interaction.

When Eq. (\ref{MainEq}) is written in the non-relativistic
approximation for the RGPEP, spin, and integration measure
factors, the limit of small coupling exhibits scaling
property similar to the Schr\"odinger equation for
positronium. The scaling in RGPEP is described at the
beginning of Appendix \ref{RGPEPscaling}. The scaling
implies that the quark eigenvalue problem is dominated by
the relative momenta on the order of strong Bohr momentum, 
\begin{eqnarray}
\label{kB}
k_B & = & { 4 \over 3} \, \alpha_0 
\, { m \over 2 } \, ,
\end{eqnarray}
which is the quark analog of $\alpha \, m_e/2$ in
positronium in QED. This scale emerges from the window
eigenvalue condition independently of the value of
$\lambda_0$ as long as $\lambda_0$ is sufficiently large in
comparison to $k_B$. In other words, the eigenvalues $M^2$
depend mainly on the value of $\alpha_{\lambda_0}$ and not
on the value of $\lambda_0$ itself when $\lambda_0$ is much
larger than $k_B$. The questions of how large $\lambda_0$
must be in comparison to $k_B$ in order to obtain results
that are sensitive to $\lambda_0$ practically only through
the value of $\alpha_0$, or to what extent this lack of
direct sensitivity to $\lambda_0$ itself is obtained for
realistic values of $\alpha_0$ and $\lambda_0$, are not
answered here. A study of such issues has been done before
in a model based on Yukawa theory
\cite{largep,MWieckowskiPhD}.

In a formal analysis of the non-relativistic expansion
for infinitesimal coupling constant $\alpha_0$, one can 
assume 
\begin{eqnarray}
\label{lambda0}
\lambda_0 
& = & 
\lambda_m \, {m \over 2}  \, , \\
\label{lambdam}
\lambda_m 
& = & 
\left( { 4 \over 3} \, \alpha_0 \right)^{0.5 + \epsilon} \,
\lambda_p \, .
\end{eqnarray}
Thus, $\lambda_0$ is much smaller than $m/2$ when
$\lambda_p$ is on the order of 1. The two parameters
$\lambda_p$ and $\epsilon$ are useful in separating
different terms in the complex, spin-dependent
interactions that otherwise do not occur clearly
ordered in size. The particular choice of the power 0.5
+ $\epsilon$ follows from how the form factors $f$
limit momentum transfers in vertices. The most
interesting case is $\epsilon$ close to 0 (see below).
At the same time, $k_B$ is much smaller than
$\lambda_0$ as long as $0 < \epsilon < 0.5$. In the
formal analysis, $k_B$ is considered much larger than
$\Lambda_{QCD}$ when one evaluates the window
$W_{\lambda_0}$ using RGPEP. But when one extrapolates
the coupling constant in the window to realistic
values, a realistically large value of $\Lambda_{QCD}$
is introduced, instead of an infinitesimally small one.
The ansatz for the gluon mass gap corresponds to the
scale of the realistic $\Lambda_{QCD}$.

Note that the formally introduced relationship between
$\lambda_0$ and $\alpha_0$ does not mean that one replaces
the true RGPEP dependence of $\alpha_0$ on $\lambda_0$ by an
artificial one. All that is done is to introduce a
parameterization of an unknown infinitesimal value of
$\alpha_0$ at a single value of $\lambda_0$; the parameters
$\lambda_m$, $\lambda_p$, and $\epsilon$, remain free to
change while $k_B$ stays always much smaller than
$\lambda_0$. After the scaling picture is described using
this parameterization and identifying terms that scale with
different powers of $\alpha_0$, one can look for the values
of $m$ and $\lambda_m$ for which the scaling picture
extrapolated to large values of the coupling constant is
useful phenomenologically. At that point one identifies the
realistic values of $\alpha_0$ and $\lambda_0$. All one
obtains this way is an approximate picture for heavy
quarkonium dynamics that can serve as a starting point for a
systematic calculation of corrections using RGPEP. 
%
\subsection{ Scaling of different factors }
%
The scaling expansion of RGPEP factors $A$ and $B$ that
occur in the potential $v_0 (13,24)$ in Eq. (\ref{MainEq}),
in terms of powers of $\alpha_0$ (we shall omit the
subscript $0$ below whenever it is irrelevant to the
context) is given in Eqs. (\ref{Aappendix}) and
(\ref{Bappendix}). The first terms in the scaling expansion
that provide the leading approximation are 
\begin{eqnarray}
\label{Atext} 
A & \simeq & - f \, {1 \over q^2 } \, + \, ff
\, { 1 \over q^2 } \, {\mu^2 \over q^2 + \mu^2 } \, , \\
\label{Btext} B & \simeq & 4m^2 \, ff \, { 1 \over q_z^2 }
\, {\mu^2 \over q^2 + \mu^2 } \, . 
\end{eqnarray} 
The symbol $q$ denotes the three-momentum transferred between 
quarks. The self-interaction terms $\delta m^2$ contain the 
same function $B$ in their integrands. 

The RGPEP factors $ff$ behave as
\begin{eqnarray}
ff & \simeq & 
\exp{\left[ - 2\left({mq^2 \over q_z \lambda^2}\right)^2 \right]} \, .
\end{eqnarray}
Using definitions of Eqs. (\ref{scalingsubstitution})
and (\ref{scalingp}) for momenta and (\ref{lambda0})
and (\ref{lambdam}) for $\lambda_0$, one obtains
\begin{eqnarray}
ff & \simeq & 
\exp{\left\{ - {8 p^2 \over 
\left[ \left( {4 \over 3} \, \alpha \right)^\epsilon
\lambda_p \right]^4 \cos^2\theta } \right\} } \, ,
\end{eqnarray} 
where $q_z$ was replaced by $ q \cos{\theta}$ and $\theta$
is the angle between $\vec q$ and $z$-axis. For small
$\alpha$, the form factors are not zero only for small $p$,
and, in fact, only for vanishingly small $p$ when the angle
$\theta$ between $\vec p$ and $z$-axis approaches $\pi /2$.
Unless there exists a large contribution in the region of
small $p$, especially near $\theta \sim \pi/2$, from other 
factors in the potential, the factor $ff$ is equivalent to 
zero when $\alpha$ is near 0 in the scaling analysis. But $B$ 
can be large for $\theta \rightarrow \pi/2$ due to $q_z^2$ in 
denominator. This singular behavior originates in the instantaneous 
LF potential due to gauge coupling between colored particles.
One has to find the result that survives in the limit of
small $\alpha$ in the presence of the singularity. The
factor regulating the singularity at $q_z=0$ is provided by
the ansatz function $\mu^2$. When one combines the terms
with $ff$ in the self-interactions $\delta m^2$ and in $A$
and $B$ in the gluon exchange potential, the net result is a
spherically symmetric and spin-independent harmonic
oscillator potential whose spring constant is no longer
sensitive to the mass ansatz $\mu^2$ under quite general
assumptions \cite{ho}. The oscillator frequency is 
\begin{eqnarray} 
\label{resultomega}
\omega &=& \sqrt{ {4\over 3} \, {\alpha \over \pi} } \,\, 
           \lambda \, \left( \lambda \over m \right)^2 \,
           \left( { \pi \over 1152 } \right)^{1/4} \, ,
\end{eqnarray}
and the corresponding spring constant, $k =
m\omega^2/2$, leads in the dimensionless Schr\"odinger
equation in variables $\vec p_{ij}$, defined in Eq.
(\ref{scalingsubstitution}), to the oscillator term
that scales like $\alpha^{6 \epsilon}$ and becomes
independent of $\alpha$ when $\epsilon \rightarrow 0$,
see Eqs. (\ref{MainEqphi}) to (\ref{kp}) below. The 
oscillator potential is independent of the quark spins.

According to Eq. (\ref{v0}), all spin effects in the leading
approximation originate from the current factors that
multiply the term $-f/q^2$ in $A$ in Eq. (\ref{Atext}). The
leading spin effects can be identified using the same
scaling analysis. In the scaling analysis for infinitesimal
$\alpha$, spin-dependent terms are $\alpha^2$ times
smaller than the spin-independent terms. Therefore, one can
also try to include corrections order $\alpha^2$ that do
not depend on spin. However, the spin-independent correction 
terms order $\alpha^4$ that emerge from the scaling expansion 
based on only 2nd-order RGPEP violate rotational symmetry and 
are expected to cancel out or get corrected when the window 
is calculated in 4th-order RGPEP. Since the spin factors have 
distinct origin (quark current factors that are specific to 
QCD) than the generic RGPEP factors (the same in all field 
theories) and momentum space integration measure (universal 
in relativistic particle physics), the structure of interaction 
kernels that one encounters in the scaling equation (in addition 
to the spin-independent harmonic oscillator term) can be written 
as (the factor $\alpha^2$ in front is not included) 
\begin{eqnarray}
\label{calV}
{\cal V} & = & 
f \, \left({4\pi \over p^2} + \alpha^2 R \right)
(1 + \alpha^2 S)(1 + \alpha^2 M) \, ,
\end{eqnarray} 
where $R$ refers to the RGPEP factors, $S$ to
spin, and $M$ to measure. 

We drop the term $R$ because it depends on the $z$-axis and
can only be corrected in the 4th order calculation for RGPEP
factors. In the 2nd-order RGPEP factors analyzed here, the
correction $R$ is given by the term $c$ in Eq.
(\ref{Aappendix}) and when one averages this term over the
direction of the $z$-axis, it vanishes. Nevertheless, one
should remember that a genuine 4th-order calculation of
$H_\lambda$ may produce corrections of the type $R$ that
will change the radial dependence of the potential from the
Coulomb shape to a different one.

In order to identify the leading spin effects, we
combine the spin and measure corrections to the factor
$1 + \alpha^2(S+M)$ and write it shortly as $1 + BF$,
where $BF$ stands for Breit-Fermi terms. The point is
that one can observe cancellation between $S$ and $M$
factors and the remaining terms produce a rotationally
symmetric spin-dependent terms after one introduces the
additional turn in spin basis that is described in
Appendix \ref{additionalturn}.
%
\subsection{ Structure of the eigenvalue problem }
\label{eqstructure}
Finally, using dimensionless variables $\vec p_{13}$
and $\vec p_{24}$ (Fig. \ref{fig:oge} illustrates the
labeling of the momentum variables) defined in Eqs.
(\ref{scalingsubstitution}) and (\ref{scalingp}), one
arrives at the following eigenvalue equation for the
spin-dependent $2 \times 2$ matrix wave function
$\phi$, defined in Eqs. (\ref{phi1}) and (\ref{phi2}):
\begin{eqnarray}
\label{MainEqphi}
0 & = & \left[ \vec p_{13}^{\, 2} - k_p \,
\Delta_{p_{13}} -x \right] \phi(\vec p_{13})  
- 
2 \int {d^3 p_{24} \over ( 2 \pi )^3 } \, 
{\cal V} \, \phi
(\vec p_{24}) \, , \nonumber \\
& & \\
\label{BFinV}
{\cal V} & = & f \, { 4 \pi \over p^2} \, (1 + BF) \, , \\
\label{fp}
f
& = & 
\exp{\left\{ - \left[ \left( {4 \over 3} \alpha \right)^{1 - 2 \epsilon} \,
     { p_{13}^2 - p_{24}^2 \over \lambda_p^2 }
\right]^2 \right\} } \, , \\
\label{kp}
k_p 
& = &
{1 \over \sqrt{ 1152 \, \pi}} \, 
{\lambda_p^6 \over 16} \, \left( {4 \over 3} \alpha \right)^{6\epsilon} \, .
\end{eqnarray}
In the limit $\alpha \rightarrow 0$
in the above result, the eigenvalues $x$ tend to $
-1/n^2$ with natural $n$ (the Coulomb spectrum). One
obtains meson masses by evaluating the eigenvalues 
$x$ for realistic values of $\alpha$ and using the 
formula 
\begin{eqnarray} 
\label{Massx} 
M & = & 2m \, \sqrt{ 1 +
x \, \left( {2\over 3} \alpha \right)^2 } \, .
\end{eqnarray} 

The fact that a mass-gap ansatz for gluons leads to an
oscillator-like interaction term, which respects rotational
symmetry already in the 2nd-order analysis, in which $\mu^2
\sim 1$, does not seem accidental. The result is almost
independent of all details of the ansatz because $q^2$ is
limited by the form factors $ff$ in the function $B$ of Eq.
(\ref{Btext}) to so small values that the ratio $\mu^2/(q^2
+ \mu^2)$ is practically 1 for any reasonable ansatz. In
addition, it seems not excluded that the same result comes
out also as a part of the genuine 4th-order calculation. In
the 4th-order calculation, the part of the ansatz for
$\mu^2$ that is order $1$ may cancel out in the window $W$
with large $\alpha$. Looking at Eqs. (\ref{ansatzscheme2})
and (\ref{ansatzscheme2}), one can see that the ansatz term
order 1 in the three-body sector is eliminated in 4th-order
RGPEP calculation when one includes the three-body sector in
the non-perturbative window dynamics. If instead one uses
the perturbative operation $R$ to further reduce the window
to the two-body quark-antiquark sector only, the cancellation
of the ansatz begins in 6th-order calculation. For large 
couplings, there can exist cancellations that cannot be easily
foreseen. But even if the ansatz is not introduced at all,
some shift of the three-body invariant mass, say $\mu^2_1$,
will emerge from QCD interactions of formal order $\alpha$
in the 3-body sector and this is how the actual gap may show
up for the first time. When one proceeds to the scaling
analysis of functions analogous to $A$ and $B$ in Eqs.
(\ref{Atext}) and (\ref{Btext}), the new shift will be of
order $\alpha$ if it is proportional to $m^2$, of order
$\alpha^{3/2}$ if it is proportional to $m \lambda$, and of
order $\alpha^2$ if it is proportional to $\lambda^2$. But
the momentum transfer squared, $q^2$, is of order $\alpha^2$
and it may continue to be formally much smaller than or
comparable to $\mu^2_1$ in the scaling limit of small
$\alpha$. The form factor $f$ limits $|\vec q \,|$ to values
on the order of $(|t|/2)\, \lambda_p^2 \, (4\alpha/3)^{2
\epsilon} \, k_B $, where $t = \cos{\theta}$ and $\theta$ is
the angle between the momentum of the effective gluon and
the $z$-axis. The terms that lead to the harmonic force
originate from the singular behavior of $q_z^{-2} \sim
1/t^2$ when $t \rightarrow 0$. But one can still neglect
$q^2 \sim t^2 \alpha^ {2+4\epsilon}$ in comparison to
$\mu^2_1 \sim \alpha^n t^{1 + \delta_\mu}$ with $n=1$ and
3/2, and even for $n=2$ the result of integration may be
close to the one obtainable when $q^2$ is neglected in
comparison to $\mu^2_1$, cf. \cite{ho}. Also, if the
effective mass ansatz is just a first term in the expansion
of the gluon gap in powers of the gluon momentum squared,
which corresponds to the case with $\mu^2 \sim q^2 t^\delta$
in the limit of small $t$ \cite{ho}, the scaling applies in
the same way and leads to the same oscillator result
\cite{ho}. So, if some mass gap shows up in order $\alpha$
in the 3-body sector, as one expects it to happen in QCD,
the results of Eqs. (\ref{Atext}) and (\ref{Btext}) may
still be valid. 
%
\section{ Spin effects }
%
\label{spin}  
Spin effects are caused by the Breit-Fermi terms, $BF$ in Eq.
(\ref{BFinV}), that originate from the product of currents
$j_{12}^\mu \bar j_{43 \, \mu}$ in $v_0$ in Eq. (\ref{v0}).
In the leading approximation, $v_0$ is displayed in Eq.
(\ref{v0approximate}). The spinors $u_{k_i, s_i}$ and
$v_{k_j, s_j}$ in the currents originate from the canonical
LF Hamiltonian of QCD and they are related by boosts ${\cal
L}_{ij}$ described in Appendix \ref{additionalturn} to the
spinors $u_{\vec k_{ij}, s_i}$ and $v_{-\vec k_{ij}, s_j}$
that are introduced in the definition of the CMF wave
function $\Psi_{CMFij}(\vec k_{ij})$ in Eq.
(\ref{wavefunctionCMF}). 

It is shown in Appendix \ref{BFterms} that the 
BF terms in Eq. (\ref{BFinV}) are
\begin{eqnarray}
\label{BFtext}
BF \phi 
& = & 
{ \alpha^2  \over 9} \,
\left[ 3( p_{24}^2 + p_{13}^2 ) \, a  
- 
\vec p \, \vec \sigma \,
\vec b \, \vec \sigma \,
\vec p \, \vec \sigma 
\right. \nonumber \\
& + & 
\left.
3 \, \vec p_{13} \vec \sigma \,\, \vec p_{24} \vec \sigma  \,\, \vec b \, \vec \sigma 
 +  
3 \, \vec b \, \vec \sigma \,\, \vec p_{24} \vec \sigma \,\, \vec p_{13} \vec \sigma  
\right] \, ,  
\end{eqnarray}
where $\phi$ denotes the $2 \times 2$ matrix 
wave function $\phi = a + \vec b \, \vec \sigma $
and $\vec p$ is the difference between $\vec p_{13}$
and $\vec p_{24}$. An alternative form of the same result,
\begin{eqnarray}
\label{BFnotation}
BF \phi 
& = & 
{\alpha^2 \over 3} \, ( p_{24}^2 + p_{13}^2 ) \, a \nonumber \\
& + & 
{\alpha^2 \over 9} \,
\left[
( 4 \, \vec p_{13} \vec p_{24} + p_{13}^2 + p_{24}^2 )
\,\, \vec b \, \vec \sigma \right.  \\
& + &
\left.
\vec b \, (8\vec p_{24}- 2\vec p_{13} ) \,\, \vec p_{13} \vec \sigma 
-  
\vec b \, ( 4 \vec p_{13} + 2 \vec p_{24} ) \,\, \vec p_{24} \vec \sigma   
\right],  \nonumber 
\end{eqnarray}
shows that the singlet wave function $a$ and the 
triplet wave function $\vec b$ are not coupled
and describe different eigenstates.

The resulting eigenvalue equations for different mesons are
listed in subsections below. These equations are
boost-invariant and describe the relative motion of two
heavy quarks no matter how fast the whole quarkonium is
moving, which is also reflected in Eq. (\ref{Massx}) that
differs from a non-relativistic expression for energy of a
slowly moving object, $E = M + E_B + P^2/(2M)$, where the
binding energy $E_B$ is given by some Schr\"odinger
equation. In order to obtain a state of a moving quarkonium,
one has to insert the wave function $\phi$ into Eq.
(\ref{state}), using Eqs. (\ref{wavefunction}),
(\ref{wavefunctionCMF}), (\ref{phi1}), and (\ref{phi2}), all
of which are relativistic. Note that the eigenvalue
equations in the subsections below lead to non-local
interactions at short distances if one introduces position
variables canonically coupled with the relative
three-momenta of quarks defined here in a boost invariant
way. The reason for nonlocality is that the interactions
contain the form factors, $f$, that depend on the
differences of invariant masses before and after an
interaction. One should remember that although the equations
in the next subsection look deceptively simple and similar
to non-relativistic models, they describe wave functions that
from the models' point of view formally correspond to
different frames of reference for different values of the
modulus of the relative three-momentum of quarks, see Appendices.

Note also that the interaction terms in the equations listed
below include potentials that in the absence of the form
factors $f$ would produce three-dimensional Dirac
$\delta$-functions in the position space representation. The
$\delta$-functions would lead to ultraviolet divergences.
But the form factors $f$ smear the $\delta$-functions and
render finite results. Nevertheless, the terms with the
smeared $\delta$-functions are not weak when $\alpha$ is
extrapolated to values on the order of 1 and they contribute
to significant spin effects for states that involve
significant $s$-wave components. 
%
\subsection{ Mesons $\eta_c$ and $\eta_b$ }
%
In this case, $\vec b = 0$ and the eigenvalue equation for 
$\phi(\vec p_{ij}) = a_{ij}/p_{ij} $ takes the form
\begin{eqnarray}
\label{MainEqeta}
0 & = & \left[ \vec p_{13}^{\, 2} - k_p \,
\Delta_{p_{13}} -x \right] {a_{13} \over 2 p_{13}} 
 - 
\int {d^3 p_{24} \over ( 2 \pi )^3 } \, 
{\cal V} \, { a_{24} \over p_{24}} \, , 
\end{eqnarray} 
where 
\begin{eqnarray}
{\cal V} & = & f \, { 4 \pi \over p^2} \, 
\left[ 1 + {\alpha^2 \over 3} \, ( p_{24}^2 + p_{13}^2)\right] \, .
\end{eqnarray}
The orbital angular momentum is zero and the 
integration over angles (see Appendix \ref{angularintegrals})
produces a one-dimensional integral equation 
\begin{eqnarray}
\label{MainEqeta1}
0 & = & h_{sosc} \, a_{13}
 - 
{2 \over \pi} \int_0^\infty dp_{24} \, f \, p_{13} p_{24} \, {\cal W} \, a_{24} \, ,
\end{eqnarray} 
with
\begin{eqnarray}
{\cal W} & = & 
\left[ 1 + {\alpha^2 \over 3} \, ( p_{24}^2 + p_{13}^2)\right] \, J_0 \, ,
\end{eqnarray} 
where the function $J_0$ is given in Appendix \ref{angularintegrals}. 
\begin{eqnarray}
 h_{sosc} & = & p_{13}^2 - k_p \partial_{13}^2 - x  
\end{eqnarray}
is introduced as a generic notation for the $s$-wave
harmonic oscillator terms in the mass eigenvalue 
equations for all mesons.
%
\subsection{ Mesons $J/\Psi$ and $\Upsilon$ }
%
In this case $a = 0$ and the eigenvalue equation 
describes a function $\phi(\vec p_{ij}) = \vec b_{ij} 
\vec \sigma $, where 
\begin{eqnarray}
\label{Upsilon}
b_{13}^k
& = & 
\left[ \delta^{kl} \, { S_{13} \over p_{13}} + 
{1 \over \sqrt{2}} \, 
\left( \delta^{kl} - 3  { p_{13}^k p_{13}^l \over p_{13}^2 } \right) \, 
{ D_{13}  \over p_{13}} \right] \, s^l \, , \nonumber \\ 
\end{eqnarray}
and $\vec s$ is a polarization vector of a massive meson
of spin 1. The $s$-wave wave function $S$ and $d$-wave wave 
function $D$ satisfy two coupled equations
\begin{eqnarray}
0 & = & \left[
\begin{array}{cc}
h_{sosc} , & 0  \\
0, & h_{sosc} + k_p {6 \over p_{13}^2} 
\end{array}
\right]
\,
\left[
\begin{array}{c}
S_{13}  \\
D_{13}
\end{array}
\right] \nonumber \\
& - &
{2 \over \pi} \int_0^\infty dp_{24} \, f \,  p_{13} p_{24}
\left[
\begin{array}{cc}
{\cal W}_{ss} , &  {\cal W}_{sd} \\
{\cal W}_{ds} , &  {\cal W}_{dd}
\end{array}
\right]
\,
\left[
\begin{array}{c}
S_{24}  \\
D_{24}
\end{array} 
\right] \, , \nonumber \\
\end{eqnarray}
where
\begin{eqnarray}
{\cal W}_{ss} & = & J_0  + {\alpha^2 \over 3} 
\left[  (p_{13}^2 + p_{24}^2) \, J_0 - 16/9\right] \, ,  \\
{\cal W}_{sd} & = &  
{\alpha^2 \over 3} \left[ p_{13}^2 \, (J_2-J_0) +  4/3 \right] {\sqrt{2}\over 3} \, , \\ 
{\cal W}_{ds} & = &  
{\alpha^2 \over 3} \left[ p_{24}^2 \, (J_2-J_0) +  4/3 \right] {\sqrt{2}\over 3} \, , \\ 
{\cal W}_{dd} & = & 
J_2  +  (J_2 - J_0)/2  \\
& + & 
{\alpha^2 \over 3} 
\left\{ (p_{13}^2 + p_{24}^2) \, \left[J_0 - (J_2-J_0)/6 \right] - 20/9  \right\} \, , \nonumber 
\end{eqnarray}
and the functions $J_0$ and $J_2$ are given 
in Appendix \ref{angularintegrals}. 
%
\subsection{ Mesons $\chi_{c_0}$ and $\chi_{b_0}$ }
%
Here $a = 0$ and the eigenvalue equation for 
$\phi(\vec p_{ij}) = b_{ij} \, \vec p_{ij} \, 
\vec \sigma /p_{ij}^2 $ takes the form 
\begin{eqnarray}
\label{MainEqchi0}
0 
& = &
\left( h_{sosc} + k_p {2 \over p_{13}^2 } \right) \, b_{13} 
 - 
{2 \over \pi} \int_0^\infty dp_{24} \,  
f \, p_{13} p_{24} \, {\cal W} \,  b_{24} \, , \nonumber \\
\end{eqnarray}
where (see Appendix \ref{angularintegrals}) 
\begin{eqnarray}
{\cal W} 
& = &
J_1 +  
{\alpha^2 \over 9} \, 
\left[ p_{13} p_{24} \, 8J_0 - \left( p_{13}^2 + p_{24}^2 \right)  J_1 \right] \, .
\end{eqnarray}
%
\subsection{ Mesons $\chi_{c_1}$ or $\chi_{b_1}$ }
%
Here again $a = 0$ and the radial eigenvalue equation for 
$\phi(\vec p_{ij}) = b_{ij} \, \vec s \times \vec p_{ij} \, 
\vec \sigma / p_{ij}^2 $ takes the form 
\begin{eqnarray}
\label{MainEqchi1}
0 
& = &
\left( h_{sosc} + k_p {2 \over p_{13}^2} \right) \, b_{13}  \, 
 - 
{ 2 \over \pi } \int_0^\infty  dp_{24} \, f \, p_{13} p_{24} \, {\cal W} b_{24} \, , \nonumber \\
\end{eqnarray}
where
\begin{eqnarray}
{\cal W}
& = &
J_1 + {\alpha^2 \over 9} \left[ 2 p_{13} p_{24} (J_0 + J_2)  
+ 
\left( p_{13}^2 + p_{24}^2 \right) J_1 \right] \, . \nonumber \\
\end{eqnarray}
%
\subsection{ Mesons $\chi_{c_2}$ or $\chi_{b_2}$ }
%
In this case also $a = 0$ and the eigenvalue equation
is for 
\begin{eqnarray}
\phi_{13} 
& = & {\cal S}_n \,
{p_{13}^i \over p_{13}} \, 
s_n^{ij} \, \sigma^j \,\, {P_{13} \over p_{13}}   \nonumber \\
& + &
{\cal S}_n \sqrt{ 2 \over 7} \, {p_{13}^i \over p_{13}} \, 
\left[ s_n^{ij} - {5 \over 2} \, \delta^{ij} \, 
{p_{13}^k \over p_{13}} s_n^{kl} {p_{13}^l \over p_{13}} \right]\, 
\sigma^j \,\, {F_{13} \over p_{13}}  \, , \nonumber \\
\end{eqnarray}
where $s_n^{ij}$ with $n$ = 1, 2, ..., 5 are symmetric traceless 
$3 \times 3$ matrices, ${\cal S}_n$ is the corresponding polarization 
five-vector for a massive meson with spin 2 (the sum over $n$ from 
1 to 5 is indicated only by the repeated subscript $n$). The 
$p$-wave wave function $P_{13}$ and the $f$-wave wave function $F_{13}$ 
satisfy two coupled equations
\begin{eqnarray}
\label{MainEqchi2}
0 & = & \left[
\begin{array}{cc}
h_{sosc} + k_p {2 \over p_{13}^2} , & 0  \\
0, & h_{sosc} + k_p {12 \over p_{13}^2} 
\end{array}
\right]
\,
\left[
\begin{array}{c}
P_{13}  \\
F_{13}
\end{array}
\right] \nonumber \\
& - &
{2 \over \pi} \int_0^\infty dp_{24} \, f \,  p_{13} p_{24}
\left[
\begin{array}{cc}
{\cal W}_{pp} , &  {\cal W}_{pf} \\
{\cal W}_{fp} , &  {\cal W}_{ff}
\end{array}
\right]
\left[
\begin{array}{c}
P_{24}  \\
F_{24}
\end{array} 
\right] \, , \nonumber \\
\end{eqnarray}
where
\begin{eqnarray}
{\cal W}_{pp} & = & 
J_1 
+ 
\alpha^2 \, 
p_{13} p_{24} \, { 14 \over 45} (3 J_2 - J_0)  \nonumber \\
& + &
\alpha^2 \, 
( p_{24}^2 + p_{13}^2 ) \, {1 \over 45} J_1
 \, ,  \\
{\cal W}_{pf} & = &  
- \alpha^2 \,
p_{13} p_{24} \, {2 \sqrt{6} \over 45} \, (3 J_2 - J_0) \nonumber \\
& + &
\alpha^2 \,  {\sqrt{6} \over 45}
\left[ p_{24}^2 \, 2J_1 
+
p_{13}^2 \, ( 5J_3 - 3J_1 ) \right] \, , \\ 
{\cal W}_{fp} & = &  
- \alpha^2 \, 
p_{24} p_{13} \, {2 \sqrt{6} \over 45} \, (3 J_2 - J_0) \nonumber \\
& + &
\alpha^2 \, { \sqrt{6} \over 45} 
\left[
p_{13}^2 \, 2J_1 
+
p_{24}^2 \, \left( 5J_3 - 3J_1 \right) \right] \, , \\ 
{\cal W}_{ff} & = &  
(5 J_3 - 3 J_1)/2
+   
\alpha^2 \, 
p_{13} p_{24} \, { 16 \over 45} (3 J_2 - J_0) \nonumber \\
& - &
\alpha^2 \, 
(p_{13}^2 + p_{24}^2) \, {1 \over 90} ( 5J_3 - 3 J_1 ) 
\, ,
\end{eqnarray}
and the functions $J_0$, $J_1$, $J_2$, and $J_3$, are given 
in Appendix \ref{angularintegrals}. 
%
\subsection{ Singlets with $J=2$, or $^1 D_2$ }
%
In this case $\vec b_{ij} = 0$ and 
\begin{eqnarray}
\phi_{13} & = & {\cal S}_n \, { a_{13} \over p_{13} } \, { p_{13}^i s^{ij}_n p_{13}^j \over p_{13}^2} \, .
\end{eqnarray}
The eigenvalue equation for the function $a$ reads
\begin{eqnarray}
\label{MainEqJ2D1}
0 & = & 
\left[ h_{sosc} + k_p {6 \over p_{13}^2 } \right] \, a_{13} 
 - 
{2 \over \pi }
 \int_0^\infty  dp_{24} \, f \, p_{13} p_{24} \, {\cal W} \, a_{24} \, , \nonumber \\  
\end{eqnarray}
where
\begin{eqnarray}
{\cal W}
& = & 
\left[ 1 + {\alpha^2 \over 3} \, ( p_{24}^2 + p_{13}^2)\right] \, 
\left( {3 \over 2} J_2 - {1 \over 2} J_0 \right) \  .  
\end{eqnarray}
%
\subsection{ Triplets with $J=2$, or $^3 D_2$ }
%
In this case $a=0$ and 
\begin{eqnarray}
\phi_{13} 
& = & 
{\cal S}_n 
{ b_{13}  \over p_{13} } \, { p_{13}^i s^{ij}_n \, 
(\vec p_{13} \times  \vec \sigma \, )^j  \over p_{13}^2 } \, .
\end{eqnarray}
The eigenvalue equation takes the form
\begin{eqnarray}
\label{MainEqJ2D3}
0 
& = &
\left[ h_{sosc} + k_p {6 \over p_{13}^2 } \right] \, b_{13} 
- 
{2 \over \pi} 
\int dp_{24} \, f \, p_{13} p_{24} \, {\cal W} \, b_{24} \, ,  \nonumber \\
\end{eqnarray}
with
\begin{eqnarray}
{\cal W}
& = & 
\left[ 1  + {\alpha^2 \over 9} ( p_{13}^2 + p_{24}^2 )  \right] \, \left( {3 \over 2} J_2 - {1 \over 2} J_0 \right)  \nonumber \\
& + &
{4 \alpha^2 \over 9} \,  p_{13} p_{24} \, J_3  \, .
\end{eqnarray}
%
\section{ Masses and wave functions }
\label{masses}
This section describes examples that illustrate to what 
extent the simplest version of the RGPEP approach can 
reproduce masses of the known $b \bar b$ and $c \bar c$ 
mesons and how the corresponding wave functions may depend
on the relative momentum of the quarks.
\subsection{ Coupling constant and quark mass }
One potentially valid way to determine the coupling constant
$\alpha_\lambda$ and quark mass $m_\lambda$ in $H_\lambda$
at $\lambda = \lambda_0$ is to evolve their values as
functions of $\lambda$ using RGPEP from the region of large
$\lambda$, say $\lambda = \lambda_1$, where their values may
be adjusted to observables that are minimally sensitive to
the non-perturbative mechanism of binding of quarks and
gluons. For such observables, the adjustment could be made
using a perturbative expansion for the $S$-matrix for quarks
and gluons using $H_{\lambda_1}$ in the femtouniverse
\cite{Bjorkenfemtouniverse}. Although a precisely defined
calculation including bound states as asymptotic states does
not exist yet in the RGPEP approach to QCD, some patterns
expected to occur in such calculation have already been
studied. For example, the RGPEP evolution that starts at 
$\lambda_1$ must be extended down to $\lambda_0$ comparable 
to the meson mass and to reach that far one has to deal with
issues of convergence that require optimization of details 
of the method \cite{modelafbs3}. It is also known \cite{glambda} 
that the differential equations of RGPEP that describe the 
evolution of the operator $H_\lambda$ (not the $S$-matrix) 
produce in third-order perturbation theory in QCD the coupling 
constant that evolves with $\lambda$ according to the formula 
\begin{eqnarray}
\label{alpha}
\alpha_0 = { \alpha_{M_Z} \over 1 + [\alpha_{M_Z}/( 6 \pi)] \, 
(11 N_C - 2 n_f) \ln {(\lambda_0/M_Z)} } \, ,
\end{eqnarray}
which matches the well-known formula for the running coupling
constant in the original Lagrangian calculus for the QCD action
\cite{af1,af2}. Here, $\alpha_{M_Z}$ is the coupling constant in
$H_{\lambda_1}$ with $\lambda_1 = M_Z$, $M_Z=92.1$ GeV is the
mass of the $Z$-boson, $N_C=3$ is the number of colors and $n_f$
is the number of flavors (the theory analyzed here has $n_f=1$). 
Thus, as soon as one estimates the value of the coupling constant 
$\alpha$ at one value of $\lambda$ in the boost-invariant Hamiltonian 
approach, such as $\lambda_1=M_Z$, the size of $\alpha$ at other 
values of $\lambda$ is in principle known in the entire region, 
in which the perturbative RGPEP calculus for the Hamiltonians 
can be accurate. 

For example, if one assumes that $\alpha_{M_Z} \sim 0.12$
\cite{PDG}, Eq. (\ref{alpha}) produces $\alpha_0 \sim 0.326$ at
$\lambda_0 \sim 3.7$ GeV (for $n_f=6$, one would obtain $\alpha_0
\sim 0.21$). However, if one uses the same formula in a strict
expansion in powers of $\alpha_{M_Z}$ to third order only, one
obtains a result that is about 30\% smaller than $0.326$. The
reason is that Eq. (\ref{alpha}) predicts an increasingly rapid
growth of the coupling constant when $\lambda$ decreases. But the
perturbative formula is replaced by a non-perturbative one for
$\lambda$ near the bound-state mass, and this can be studied in
detail using models \cite{modelafbs1,modelafbs2}. By analogy with
the models, one may expect a finite but rapid transition of
$\alpha$ from the increasing to a decreasing function of
$\lambda$ in the region where the binding mechanism dominates
dynamics, smoothing the discontinuity present in the perturbative
formula (\ref{alpha}) for $\alpha$ when its denominator passes
through zero. Therefore, a low-order expansion in powers of
$\alpha_{M_Z}$ is not useful. But it is useful to expand the
operator $H_{\lambda_0}$ in powers of $\alpha_0$. The operator
coefficients of this expansion can be found assuming that
$\alpha_0$ is infinitesimally small \cite{modelafbs3}. The same
coefficients can be used for evaluating $H_{\lambda_0}$ when
$\lambda_0$ is small and $\alpha_0$ is comparable with 1. Finding
a precise formula for $\alpha_0$ in terms of $\alpha_{\lambda_1}$
for realistic values of the coupling constants may require
sophisticated high-order RGPEP calculations. If the precise value
of $\alpha_0$ cannot be easily calculated in a low-order
perturbation theory, one can seek values of $\alpha_0$ that may
correspond to the available bound-state data and then incorporate
the resulting picture in a new perturbation theory around the
first approximation found that way. At the present stage of
development, one can only verify if the simplest version of the
boost-invariant Hamiltonian approach can reproduce known masses
of heavy quarkonia when the coupling constant $\alpha_0$ is
allowed to take values on the order of 1. 

Less is known about the quark mass $m$ as a function of
$\lambda$, and what values of $m_{\lambda_0}$ one should
expect in $H_{\lambda_0}$. Technically, the mass parameter
is specified as the perturbative eigenvalue of
$H_{\lambda_0}$ for one-quark states~\cite{ho}. Therefore,
one can expect that the mass should be close to the pole
mass \cite{PDG}, which is about 10\% larger than the quark
mass in the minimal subtraction scheme for bottom quarks.
One may expect for $b\bar b$ mesons that $m=m_{\lambda_0}
\sim$ 4.5 GeV to 5 GeV.

Thus, although in principle the boost-invariant Hamiltonian
approach appears able to cover the whole range of energy scales
accessible experimentally in the case of heavy quarkonia, one
needs to carry out higher-order calculations than carried out
here in order to correlate high-energy perturbative results in
the femtouniverse with the description of binding of quarks and
gluons at the scale of $b$ or $c$ quark masses in one and the
same scheme. In the simplest version of the Hamiltonian approach,
one can only find out if there exist choices for the parameters
$\alpha$ and $m$ at $\lambda_0$ on the order of the quark masses
that can produce spectra of masses of the quarkonia with
reasonable accuracy. Since one expects $\alpha \sim$ 1/3,
``reasonable'' means here that the masses should be reproducible
with accuracy on the order of 1/3 or perhaps 1/10 of the largest
splittings between states with neighboring quantum numbers. The
latter are on the order of 500 MeV and this means that matching
data with accuracy on the order of 50 MeV would be quite good in
the simplest version if the required $\alpha$ and $m$ for such
matching are close to the values established from other
considerations.

In principle, masses of only two mesons are sufficient to fix the
values of $\alpha$ and $m$ as functions of $\lambda$ near
$\lambda_0$. The question is which two masses one should
use. That the choice is not obvious and that there is a need
for a good choice is a consequence of the fact that in
approximate calculations all masses are calculated with
theoretical errors that are not known and if one uses two
masses that are obtained with a large theoretical error then
results for all other masses will be obtained with large
errors. Experience with exactly solvable models dictates
that the most accurate procedure should be to choose the
masses in the middle of the spectrum of the window
Hamiltonian \cite{largep,modelafbs3}. Let us consider the example
of $b\bar b$ mesons. The most rational choice is to use
masses of two $p$-wave mesons $\chi_1$(1P) and $\chi_1$(2P).
Their masses lie in the middle of the window spectrum. The
high-energy boundary of the window corresponds to short
distance dynamics, i.e., the most tightly bound states,
having the smallest masses. The low-energy boundary
corresponds to long distance dynamics, i.e., the states with
largest masses. The mesons $\chi_1$(1P) and $\chi_1$(2P) are
not very sensitive to the short distance dynamics and thus
also not very sensitive to the unknown term $\alpha^2 R$ in
the potential of Eq. (\ref{calV}), because quarks in these
mesons are pushed out from the region of small relative
distances by the centrifugal barrier with $l=1$. One expects
that RGPEP of 4th-order will produce terms $\alpha^2 R$ that
correspond in position space to functions like $\delta^3
(\vec r \,)$ or $1/r^3$. Such functions are known to occur
in effective potentials in standard dynamics in atomic
calculations in QED when one includes effects due to the
exchange of two photons, vertex corrections, and
self-interactions order $\alpha^2$ \cite{krp2}. Thus,
selecting mesons $\chi_1$(1P) and $\chi_1$(2P) that have
$l=1$, one has a chance to avoid theoretical errors due to
the current lack of knowledge of the terms $\alpha^2 R$ in
LF QCD. At the same time, the masses of mesons $\chi_1$(1P)
and $\chi_1$(2P) are most probably less sensitive to the
quark-antiquark long-distance dynamics than the masses of
states with $l=2$ or 3. At long inter-quark distances, the
harmonic oscillator potential is expected to lose accuracy
because the simplest approximation does not take into
account effective gluons that may be created when quarks
move far away from each other. For example, one effective
gluon could actively participate in the non-perturbative
dynamics of states with masses that exceed the middle
eigenvalues of the window Hamiltonian by more than 1 GeV,
which is an estimate of the magnitude of mass of an
effective gluon at the scale $\lambda_0$. The estimate
indicates that one should probably fit parameters $\alpha$
and $m$ in the simplest approximation to meson masses that
do not exceed the middle masses by more than about 1 GeV,
and the mesons $\chi_1$(1P) and $\chi_1$(2P) lie in this
range. 
\begin{table}[ht]
\caption{\label{tab:bottomtrial} 
Qualitative illustration of results of the simplest approximate 
approach to heavy quarkonia in the case of masses of $b \bar b$ 
mesons (in MeV). The second column is obtained using the quark 
mass $m$ = 4856.92 MeV and coupling constant $\alpha$ = 0.32595 
for $\lambda$ = 3697.67 MeV when one demands that the masses of 
mesons $\chi_1$(1P) and $\chi_1$(2P) are reproduced using an 
auxiliary interpolation procedure described in the text, which 
is employed only to increase speed of numerical estimates in this
illustration and is accurate to a few MeV. The corresponding 
oscillator parameters are $\omega$ = 182.16 MeV and $k_p$ = 
0.157722. The third column quotes experimental data with accuracy 
to 1 MeV and the fourth column displays the difference. The fifth 
column shows precise numerical results obtained from the same 
dynamical equations for the same values of $m$, $\alpha$, and 
$\lambda$, but without errors introduced by the auxiliary 
interpolation procedure.} 
\begin{ruledtabular}
\begin{tabular}{|l|l|l|r|r|}
meson           & interpolation & experiment \cite{PDG} & difference &  precise  \\
\hline 
$\Upsilon$10865 &     10725     &  10865                &   -140     &  10729.7  \\
$\Upsilon$10580 &     10464     &  10580                &   -116     &  10466.9  \\
$\Upsilon$3S    &     10382     &  10355                &     27     &  10385.2  \\
$\chi_2$2P      &     10276     &  10269                &      7     &  10278.5  \\
$\chi_1$2P      &     10256     &  10256                &      0     &  10258.0  \\
$\chi_0$2P      &     10226     &  10232                &     -6     &  10228.1  \\
$\Upsilon$2S    &     10012     &  10023                &    -11     &  10013.8  \\
$\chi_2$1P      &      9912     &   9912                &     -1     &   9913.3  \\
$\chi_1$1P      &      9893     &   9893                &      0     &   9894.2  \\
$\chi_0$1P      &      9865     &   9859                &      5     &   9865.5  \\
$\Upsilon$1S    &      9551     &   9460                &     91     &   9551.8  \\
$\eta_b$1S      &      9510     &   9300                &    210     &   9510.8 
\end{tabular}
\end{ruledtabular}
\end{table}

An example of results one obtains by fitting masses of $b
\bar b$ mesons $\chi_1$(1P) and $\chi_1$(2P) for $\lambda_0=
3697.67$ MeV is given in Table \ref{tab:bottomtrial}. The
coupling constant and mass required for obtaining the two masses
at this $\lambda_0$ are $\alpha$ = 0.32595 and $m$ = 4856.92
MeV, in a good qualitative agreement with expectations (see
Eq. (\ref{alpha}) and the discussion that follows it). The
large number of digits in these numbers is a numerical
effect due to the precision of experimental data and does
not reflect the accuracy of the Hamiltonian approach to QCD
in its simplest version with only $|Q \bar Q \rangle$ sector, 
which is presumably much worse. The value of $\lambda_0$ chosen 
in this example lies in the middle of a small range of size of 
about 200 MeV in which one can vary $\lambda_0$ and numerically 
reproduce the same known values of the two meson masses with 
accuracy better than 1 ppm by varying the parameters $\alpha$ 
and $m$ as functions of $\lambda_0$. Table \ref{tab:bottomtrial} 
shows that a whole set of masses in the middle of the window 
spectrum is also close to data when the two selected masses are. 
The masses near the edges of the window, most sensitive to the
theoretical errors of the simplest version, deviate from data 
by more than the masses in the middle of the window spectrum do, 
but the magnitude of these deviations is not absurdly large. 

Results in Table \ref{tab:bottomtrial} were obtained in a
sequence of steps that need to be explained. The key
difficulty is that the masses can be determined only
numerically, and the integrals that determine matrix
elements of the window Hamiltonians depend simultaneously
and in nontrivial way on $\alpha$, $m$, and $\lambda$. The
complication is caused mainly by the form factors $f$, which
eliminate the possibility of analytic integrations using the
oscillator basis functions. The numerical evaluation of the
matrix elements of window Hamiltonians takes time. The time
required for evaluation of one matrix element on a good
laptop is on the order of a second, and one needs on the
order of a thousand matrix elements to obtain accuracy of
four digits for masses of mesons that result from
diagonalization of the window matrix. One would have to
carry out very long computations to find suitable $\alpha$
and $m$ for any given choice of $\lambda_0$ if one were
computing matrix elements always a new for every change in
the parameters. Instead, one can evaluate eigenvalues $x$ in
Eqs. (\ref{MainEqeta1}) to (\ref{MainEqJ2D3}) using
parameters that lie on discrete points of a grid in the
parameter space. The parameter $k_p$ is more convenient than
$\lambda$ itself. Then one can interpolate between the grid
points in order to find approximate eigenvalues for a
continuum of parameters $\alpha$ and $m$ for a whole range
of values of $\lambda_0$ in the region covered by the grid.
Such interpolation produces quickly results of precision
better than $10^{-3}$. The interpolating functions allow one
to identify the values of $\alpha$ and $m$ that reproduce
the same masses of $\chi_1$(1P) and $\chi_1$(2P) for
different values of $\lambda_0$ very efficiently even though
the eigenvalues are less precise than to 1 MeV. Results of the
interpolation are given in Table \ref{tab:bottomtrial} in
the second column, marked ``interpolation.'' Precisely
evaluated masses for the parameters selected using the
interpolation are given in the fifth column in Table
\ref{tab:bottomtrial}. The quoted values were found stable 
against a) increase of the number of basis states above 
about 40 (about 20 if only one orbital angular momentum wave 
function is present), b) changes in the oscillator basis 
functions of Appendix G due to variation in the oscillator 
frequency within about one order of magnitude, and c) 
changes in the algorithm for evaluation of matrix elements 
of the window Hamiltonians (two different integration routines 
produced the same results). Subsequent discussion concerns 
results that satisfied the same convergence tests.

The example given in Table \ref{tab:bottomtrial} shows
that even in its simplest version the Hamiltonian approach can
lead to phenomenologically reasonable results for the masses of
$b \bar b$ mesons. But the example does not provide information
about how large is the range of parameters for which the simplest 
version of the Hamiltonian approach can match the masses of known 
heavy quarkonia. This issue is taken up using examples in the 
remaining parts of this section.
%
\subsection{ Masses of $b \bar b$ mesons }
%
In order to obtain qualitative information about the distance
between the simplest version of LF QCD and data for masses of $b
\bar b$ mesons, one can consider two different fits of $\alpha$
and $m$. 

\begin{figure}[t]
\includegraphics[scale=.91]{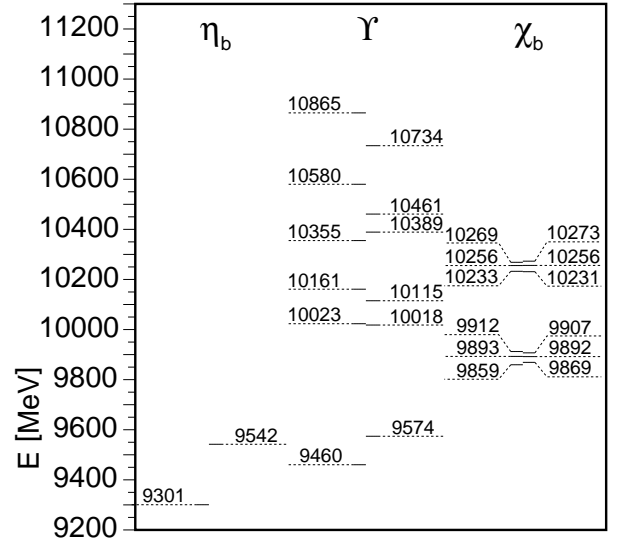}
\caption{\label{fig:bmid} Illustration of masses in 
the 3rd column in Table \ref{tab:bottom}. The left 
thick bars in each of the three columns indicate data 
and right thick bars results of computation. The theory 
mass 10461 MeV corresponds to a state dominated by the 
$d$-wave, apparently not easy to identify experimentally 
\cite{PDG}.}
\end{figure}

One fit is focused on the middle of the mass spectrum of known 
mesons. Instead of fixing two selected meson masses and checking how
others are reproduced as it was done in the previous subsection,
one finds a minimum of deviation of the computed masses from data
as function of $\alpha$ and $m$ assuming different $\lambda_0$ and
this is done for 7 masses in the middle of the experimentally known 
spectrum. Results of this fit are denoted in Table \ref{tab:bottom} 
as ``fit to middle.'' They are shown graphically in Fig.
\ref{fig:bmid}. Variation of the obtained spectrum with
$\lambda_0$ when one keeps any two of the 7 masses fixed, or
rather the degree of absence of such variation, is not further
studied in this and the next subsection. 
\begin{table}[h]
\caption{\label{tab:bottom} 
Masses of $b \bar b$ mesons (in MeV). 
The third column results from the fit of the coupling constant 
$\alpha$ and quark mass $m$ at the indicated optimal value of 
$\lambda$ to 7 middle masses of known $b \bar b$ mesons, i.e., 
masses of $\chi_0$(1P), $\chi_1$(1P), $\chi_2$(1P), $\Upsilon$(2S), 
$\chi_0$(2P), $\chi_1$(2P), $\chi_2$(2P), and this fit implies 
the oscillator parameters $\omega$ = 184.62 MeV and $k_p$ = 0.26667.
The fourth column results from the fit to masses of all 12 mesons
$\eta$(1S), $\Upsilon$(1S), $\chi_0$(1P), $\chi_1$(1P), $\chi_2$(1P), 
$\Upsilon$(2S), $\chi_0$(2P), $\chi_1$(2P), $\chi_2$(2P), 
$\Upsilon ^3D_1$(estimated as similar to $D_2$), $\Upsilon$(3S), 
$\Upsilon$ 10580(S4), and the corresponding oscillator parameters 
are $\omega$ = 147.11 MeV and $k_p$ = 0.016667. Question marks 
regarding $D$ states are explained in the text.}
\begin{ruledtabular}
\begin{tabular}{|l|l|r|r|}
                   & experiment \cite{PDG} & fit to middle &  fit to all  \\
\hline 
$\lambda$ [MeV]    & --------------        &   3779.8      &     3252.3   \\ 
$m$       [MeV]    & --------------        &   4835.9      &     4979.7   \\    
$\alpha$           & --------------        &  0.28839      &    0.50738   \\
$\Upsilon$10580    & 10580    $\pm$   3.5  &    10734      &      10629   \\ 
$\Upsilon$ $^3D_1$ & --------------        &    10461      &      10446   \\
$\Upsilon$3S       & 10355.2  $\pm$   0.5  &    10389      &      10329   \\
$\chi_2$2P         & 10268.5  $\pm$   0.72 &    10273      &      10272   \\
$\chi_1$2P         & 10255.5  $\pm$   0.72 &    10256      &      10241   \\
$\chi_0$2P         & 10232.5  $\pm$   0.9  &    10231      &      10194   \\
$\Upsilon$ $^1D_2$ & 10161.1? $\pm$   2.2  &   ------      &      10172   \\
$\Upsilon$ $^3D_2$ & 10161.1? $\pm$   2.2  &   ------      &      10169   \\
$\Upsilon$ $^1D_1$ & --------------        &    10115      &      10154   \\
$\Upsilon$2S       & 10023.3  $\pm$   0.31 &    10018      &       9991   \\ 
$\chi_2$1P         & ~9912.21 $\pm$   0.57 &     9907      &       9943   \\
$\chi_1$1P         & ~9892.78 $\pm$   0.57 &     9892      &       9908   \\
$\chi_0$1P         & ~9859.44 $\pm$   0.73 &     9869      &       9849   \\
$\Upsilon$1S       & ~9460.3  $\pm$   0.26 &     9574      &       9448   \\ 
$\eta_b$1S         & ~9300.6  $\pm$  10    &     9542      &       9359    
\end{tabular}
\end{ruledtabular}
\end{table}

\begin{figure}[b]
\includegraphics[scale=.91]{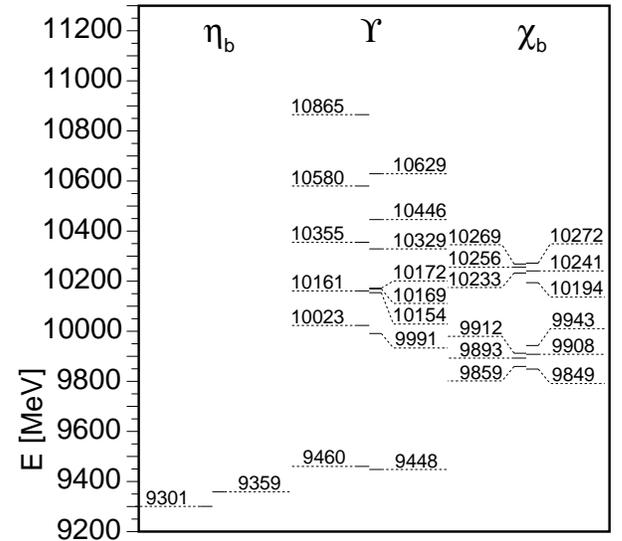}
\caption{\label{fig:ball} Illustration of masses in 
the 4th column in Table \ref{tab:bottom}. The left 
thick bars in each of the three columns indicate data 
and right thick bars results of computation. The theory 
$D$-state mass changes by only a few MeV, to 10446 MeV,
when one changes ``fit to middle'' to ``fit to all.''}
\end{figure}

The other fit is focused on checking how many of the experimentally 
known meson masses can be explained in the simplest version of LF 
QCD, and how accurately. The second fit includes masses of all 12
well-established mesons \cite{PDG} with decay widths significantly 
smaller than 100 MeV. A decay width comparable with 100 MeV is
considered an indicator of relevance of processes that the simplest 
approximate version of LF QCD cannot describe. The second fit is 
denoted in Table \ref{tab:bottom} as ``fit to all.'' The results are 
shown graphically in Fig. \ref{fig:ball}.

A separate comment is required concerning the $D$ states 
($d$-waves) in Table \ref{tab:bottom}. One such state is 
known experimentally \cite{PDG}, most probably having 
total $J$=2. This state is included in the table in order 
to illustrate what happens in the simplest version of LF
QCD concerning $d$-wave mesons. All $D$ states, whether 
one considers singlets or triplets, or $J$ = 1, are expected 
in other approaches to have similar masses \cite{Dmesons1,Dmesons2}. 
The same happens in LF QCD. In Table \ref{tab:bottom}, the 
state $\Upsilon ^1D1$ is known only in theory, and masses 
of both states $\Upsilon ^1D2$ and $\Upsilon ^3D1$ are near 
the one experimentally know mass of 10161 MeV for the 
$D$-meson that presumably has $J=2$. The excited triplet
state $\Upsilon ^3D1$ is not known experimentally but comes 
out of the calculation as dominated by its $d$-wave component. 
In Fig. \ref{fig:bmid} for $b\bar b$ mesons, the known mass 
of 10161 MeV is shown in comparison with the theoretical 
mass of the $\Upsilon$ state dominated by $d$-wave, and, 
in Fig. \ref{fig:ball}, it is shown in comparison with all 
three theoretical results for $D$ states with $J=2$ and $J=1$. 

The range of coupling constants $\alpha_0$ and quark masses $m_0$ 
for which the Hamiltonians $H_{Q \bar Q \lambda_0}$ can match the
data for $b \bar b$ mesons, as illustrated in Tables \ref{tab:bottomtrial}
and \ref{tab:bottom}, and in Figs. \ref{fig:bmid} and \ref{fig:ball}, is 
summarized using these examples in Table \ref{tab:bottom3} (subscript 0 is 
omitted).
\begin{table}[h]
\caption{\label{tab:bottom3} 
Parameters in $H_{\lambda_0}$ for $b \bar b$ mesons.} 
\begin{ruledtabular}
\begin{tabular}{|l|l|r|r|}
parameter       &  $\chi_1$  & 7 middle  &  all 12   \\
\hline 
$\lambda$ [MeV] &  3697.67   &  3779.8   &  3252.3   \\ 
$m$       [MeV] &  4856.92   &  4835.9   &  4979.7   \\    
$\alpha$        &  0.32595   &  0.28839  &  0.50738  
\end{tabular}
\end{ruledtabular}
\end{table}
Note that the fit to all 12 meson masses points to the much
larger coupling constant and quark mass at considerably smaller
$\lambda_0$ than in the two similar cases with fits to 7 or only
2 middle mesons. This feature most probably emerges because the
term $\alpha^2 R$ in the radial factor $1 + \alpha^2 R$ in 
Eq. (\ref{calV}) is set to 0 in the simplest version of the 
approach. Splittings between $s$-wave mesons, including the 
$\eta_b$ and $\Upsilon$ family, are sensitive to the short-distance 
dynamics that depends on the term $\alpha^2 R$. Calculation of 
the term $\alpha^2 R$ requires full 4th-order RGPEP analysis. 
In the absence of $\alpha^2 R$, one can nearly reproduce masses 
of the $s$-wave states at the price of increasing $\alpha$ and 
$m$. However, it is clear that the 4th-order calculation of the 
term $\alpha^2 R$ must be carried out in order to narrow the range 
of possible values of $\alpha$ and $m$.
%
\subsection{ Masses of $c \bar c$ mesons }
\begin{figure}[ht]
\includegraphics[scale=.9]{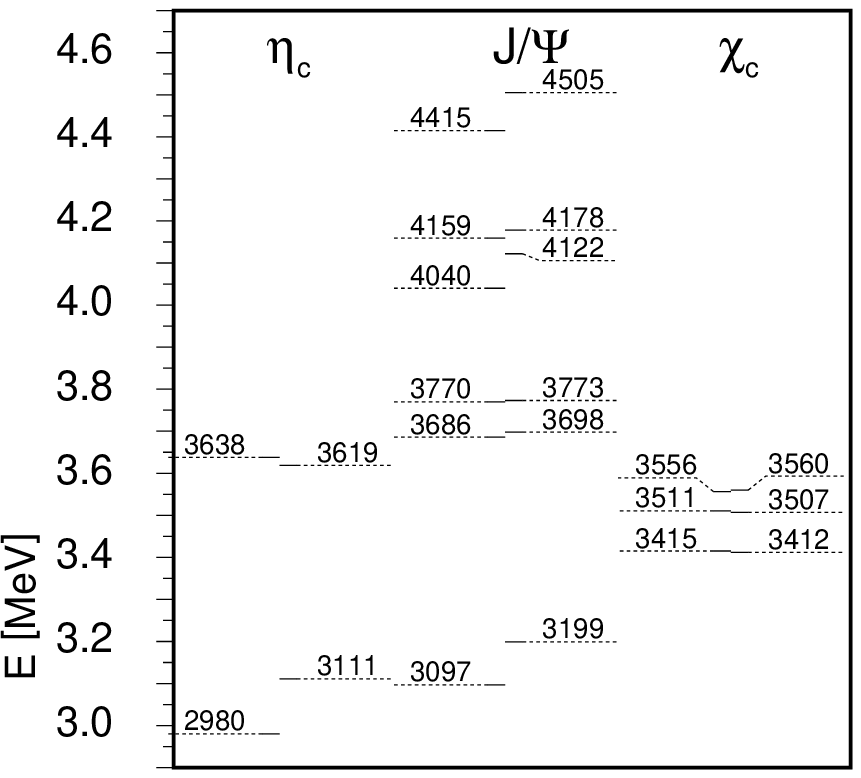}
\caption{\label{fig:cmid} Illustration of masses 
in the 3rd column in Table \ref{tab:charm}. The left 
thick bars in each of the three columns indicate data 
and right thick bars results of computation.} 
\end{figure}

\begin{figure}[hb]
\includegraphics[scale=.9]{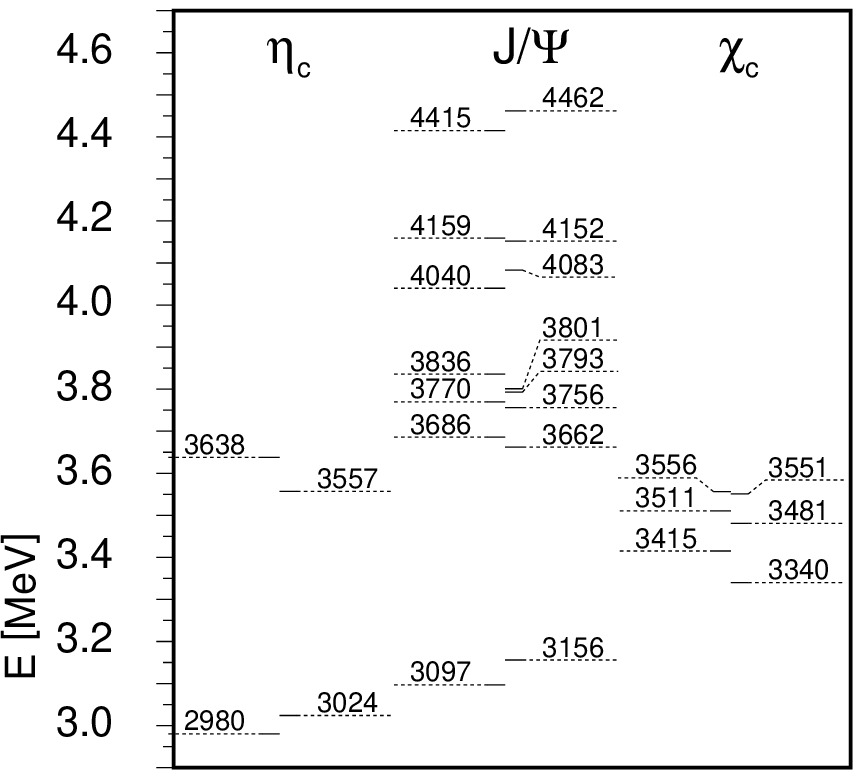}
\caption{\label{fig:call} Illustration of masses 
in the 4th column in Table \ref{tab:charm}. The left 
thick bars in each of the three columns indicate data 
and right thick bars results of computation.}
\end{figure}

%
Masses of $c \bar c$ mesons can be studied in the simplest 
version of the Hamiltonian approach analogously to the case 
of $b \bar b$ mesons discussed in the previous subsection.
Table \ref{tab:charm} shows results of two fits: to 3 middle masses
in the known spectrum, and to all 11 masses of well-established 
mesons with small decay widths. Masses of two theoretical $D$ 
states with $J=2$ can in principle be compared with the one 
experimentally known mass of 3836 MeV for a meson whose $J=2$ 
needs confirmation \cite{PDG}. Fig. \ref{fig:cmid} illustrates
the results for $c \bar c$ mesons obtained from the fit to the 
three middle masses in the window (the third column in Table 
\ref{tab:charm}), the $D$-mesons with $J=2$ are not illustrated. 
In Fig. \ref{fig:call}, illustrating results from the fit to the 
masses of all 11 $c \bar c$ mesons (the fourth column in Table 
\ref{tab:charm}), all three states with $d$-waves; the triplet 
$J=1$ state corresponding to $\psi 3770$, the triplet state, 
and the singlet state with $J=2$, are indicated. 

Table \ref{tab:charm3} shows examples of parameters that fit
masses of $c \bar c$ mesons, in comparison to the examples of
parameters from Table \ref{tab:bottom3} that fit masses of $b
\bar b$ mesons. 
\begin{table}[ht]
\caption{\label{tab:charm} 
Masses of $c \bar c$ mesons (in MeV).
The third column results from the fit of the coupling constant 
$\alpha$ and quark mass $m$ at the indicated optimal value of 
$\lambda$ to only 3 middle masses of $c \bar c$ mesons, i.e., masses 
of $\chi_0$(1P), $\chi_1$(1P), $\chi_2$(1P), and this fit implies 
the oscillator parameters $\omega$ = 284.93 MeV and $k_p$ = 3.0642.
The fourth column results from the fit to masses of all 11 mesons
$\eta$(1S), $J/\psi$(1S), $\chi_0$(1P), $\chi_1$(1P), $\chi_2$(1P), 
$\eta$(1S), $\psi$(2S), $\psi$3770, $\psi$4040, $\psi$4159, $\psi$4415, 
and the corresponding oscillator parameters are $\omega$ = 278.72 MeV
and $k_p$ = 1.3396. Question marks regarding $D$ states are explained 
in the text.}
\begin{ruledtabular}
\begin{tabular}{|l|l|r|r|}
 meson             & experiment \cite{PDG}   & fit to middle  &  fit to all \\
\hline 
$\lambda$ [MeV]    & --------------          &   1990.0       &    1934.2   \\ 
$m$       [MeV]    & --------------          &   1553.3       &    1577.4   \\    
$\alpha$           & --------------          &  0.34335       &   0.41443   \\
$\psi$4415         & 4415     $\pm$    6     &   4505         &    4462     \\
$\psi$4159         & 4159     $\pm$   20     &   4178         &    4152     \\
$\psi$4040         & 4040     $\pm$   10     &   4122         &    4083     \\ 
$^1D_2$            & 3836?    $\pm$   13     & ------         &    3801     \\ 
$^3D_2$            & 3836?    $\pm$   13     & ------         &    3793     \\
$\psi$3770         & 3770     $\pm$    2.4   &   3773         &    3756     \\
$\psi$2S           & 3686.093 $\pm$    0.034 &   3698         &    3662     \\
$\eta_c$2S         & 3638     $\pm$    5     &   3619         &    3557     \\
$\chi_2$1P         & 3556.26  $\pm$    0.11  &   3560         &    3551     \\
$\chi_1$1P         & 3510.59  $\pm$    0.1   &   3507         &    3481     \\
$\chi_0$1P         & 3415.16  $\pm$    0.35  &   3412         &    3340     \\
$J/\psi$1S         & 3096.916 $\pm$    0.011 &   3199         &    3156     \\
$\eta_c$1S         & 2980.4   $\pm$    1.2   &   3111         &    3024       
\end{tabular}
\end{ruledtabular}
\end{table}
It is plausible that the anomalously large result of $\alpha 
\sim 0.5$ for bottom quarks merely indicates that a
fit to all masses can push parameters toward explanation of the
large $s$-wave splittings at the expense of theoretical
consistency of the approach. It is expected that $\alpha$ in 
$b \bar b$ mesons should be smaller than in $c \bar c$ mesons, 
in correspondence with the increase of $\lambda_0$. Calculation 
of the term $\alpha^2 R$ in 4th-order RGPEP should clarify the
situation considerably. 

At this point, one can observe that the Hamiltonians
$H_{Q\bar Q \lambda_0}$ obtained for heavy quarkonia in the
simplest version of LF QCD seem to explain the tendency of
potential models to prefer larger values of the coupling
constant and quark masses than indicated by results based on
perturbative QCD \cite{PDG}. Namely, when one forces a single
non-relativistic Hamiltonian with a Coulomb potential at
short distances and some confining potential at large
distances to fit data for masses of many mesons, instead of
only the middle ones in the window where a simple potential 
model can be justified, the strong-interaction relativistic
effects at short distances are not well described. 
Similarly, at large distances between quarks, a simple potential 
model cannot reproduce effects due to interactions that involve 
gluons ``in the air.'' The parameters of potential models 
have to increase artificially in order to keep reproducing 
the smallest and largest masses of know mesons. 
One should mention that quark models based on the Bethe-Salpeter 
equation with a kernel that resembles a harmonic oscillator 
potential at intermediate distances have been introduced a 
long time ago~\cite{MittalMitra} and extensive studies of the 
hadronic spectrum have been made using models that successfully 
incorporate such kernels~\cite{SpenceVary}. A thorough discussion 
of potential models in a related two-body Dirac formalism is also 
available~\cite{CraterAltine}. Since the Bethe-Salpeter equation 
or two-body equations are ultimately related in QCD to an entire 
set of the Dyson-Schwinger equations~\cite{AlkoferSmekal}, one should 
observe that the whole set corresponds in the Hamiltonian approach to 
the eigenvalue problem in which all effective-particle Fock sectors 
are explicitly included.
\begin{table}[th]
\caption{\label{tab:charm3} 
Examples of parameters in $H_{\lambda_0}$ that fit masses 
of $c \bar c$ mesons, compared with the examples of parameters 
that fit masses of $b \bar b$ mesons.} 
\begin{ruledtabular}
\begin{tabular}{|l|r|r|r|r|}
parameter         & $b \bar b$ middle & $b \bar b$ all & $c \bar c$ middle & $c \bar c$ all  \\
\hline 
$\lambda_0$ [MeV] &  3779.8           &  3252.3        &   1990.0          &    1934.2       \\     
$m$       [MeV]   &  4835.9           &  4979.7        &   1553.3          &    1577.4       \\        
$\alpha$          &  0.28839          &  0.50738       &   0.34335         &   0.41443   
\end{tabular}
\end{ruledtabular}
\end{table}

A characteristic feature in Table \ref{tab:charm3} is that
the quark mass varies slowly with changes of $\lambda_0$
while the coupling constant varies relatively quickly. The
width parameter $\lambda$ occurs in third power in the
oscillator frequency, $\omega$, in Eq. (\ref{resultomega}).
Therefore, there is a possibility to keep a whole set of
meson masses approximately constant when $\lambda_0$ is
changed a little by a considerably larger percentage of
change in $\alpha$, while an a priori possible compensating
change of the quark mass cannot be large because the overall
scale of masses is dictated by Eq. (\ref{Massx}). The
eigenvalues $x$ are negative for the smallest-mass mesons
and positive for the other mesons. Thus, variation of the
quark mass is limited by the requirement of preserving the
relativistic structure of the spectrum in Eq. (\ref{Massx}).

Finally, one should stress that the adjustment of $\alpha$
and $m$ at $\lambda = \lambda_0$, which includes a choice of
$\lambda_0$, does not provide a check on the renormalization
group variation of the parameters $\alpha$ and $m$ with
$\lambda$ beyond the qualitative statement of agreement with
expectations regarding the magnitudes of the parameters. In
order to study the renormalization group structure, one
would have to consider a plausible choice of $\alpha$ and $m$ at
certain $\lambda_0$, evolve the values of $\alpha$ and $m$
in RGPEP for $H_\lambda$ to other values of $\lambda$, and
solve the eigenvalue problems for different values of
$\lambda$. One would need to include the 4th-order RGPEP to
begin with and also perform non-perturbative computations of
the spectra of window Hamiltonians with more than one Fock
sector built from effective particles. 
%
\subsection{ Wave functions }
%
\label{wavefunctions}
An important aspect of the LF Hamiltonian dynamics is that it
provides wave functions of bound states. The discussion that
follows is limited to an illustration on examples how the
Hamiltonian approach works using wave functions of mesons
$J/\psi$ and the ground state of $\Upsilon$. Masses of these
mesons are not described particularly well in the approximate
approach. But their wave functions are sufficient to display the
main features. The states corresponding to $J/\psi$ and
$\Upsilon$ contain $s$-wave and $d$-wave wave functions that are
invariant under boosts. Section \ref{states} explains how these
states are constructed using the wave functions. 

\begin{figure}[ht]
\includegraphics[scale=.4]{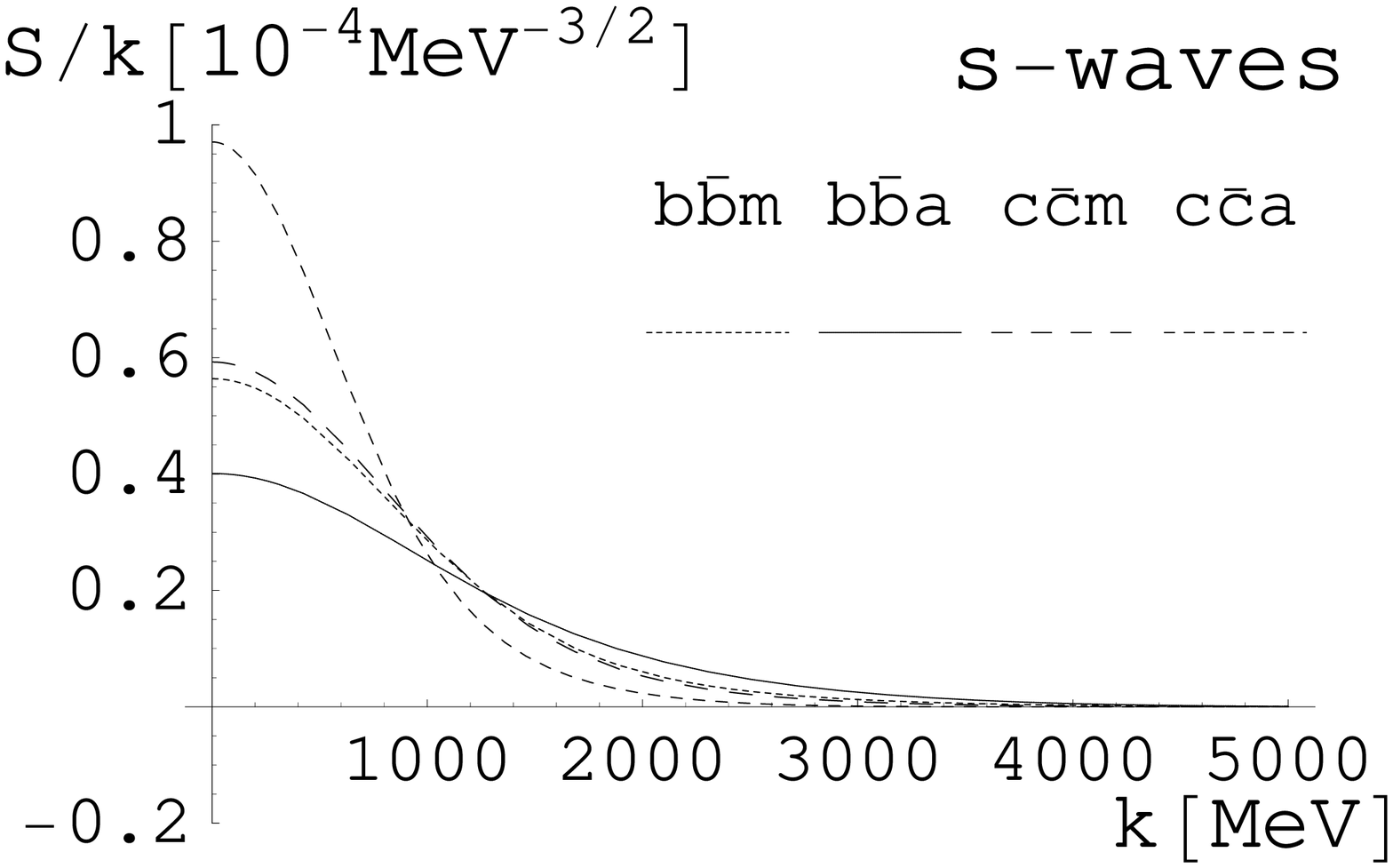}
\caption{\label{fig:swave} Four examples of $s$-wave
wave functions in $b \bar b$ and $c \bar c$ mesons. 
Two curves labeled $b \bar b$ correspond to the ground 
state of $\Upsilon$ and the two curves labeled $c \bar c$ 
correspond to $J/\psi$. The curves are presented in two 
versions that correspond to the two sets of parameters 
$\alpha$ and $m$ shown in Tables \ref{tab:bottom}
and \ref{tab:charm}. One set was adjusted to masses in
the middle of the known spectrum, and the other one to 
masses of all known mesons with small widths. The label
``m'' refers to "middle" and the label ``a'' to ``all.''}
\end{figure}
\begin{figure}[hb]
\includegraphics[scale=.4]{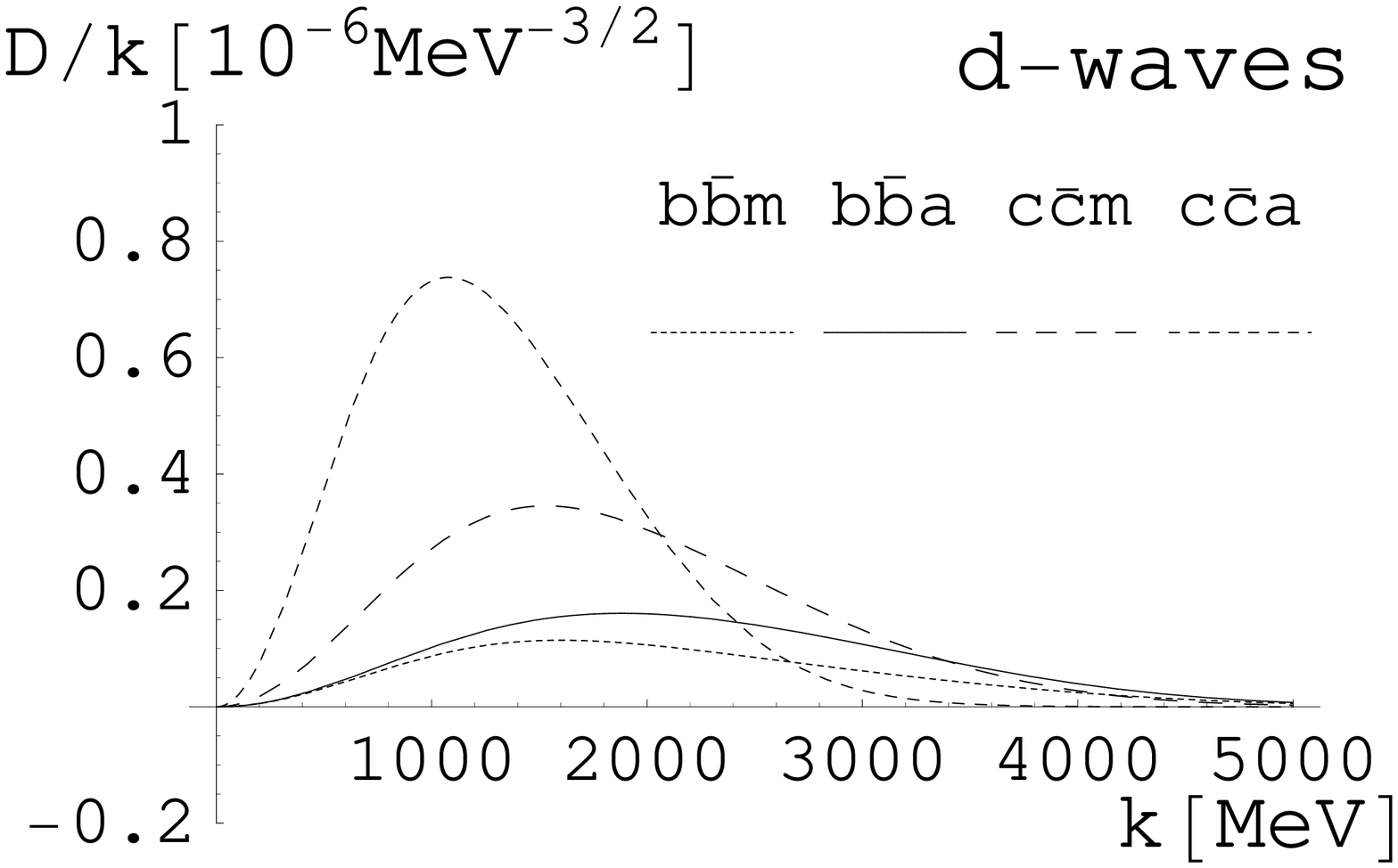}
\caption{\label{fig:dwave} Four examples of $d$-wave
wave functions in the ground state of $\Upsilon$ and 
in $J/\psi$. The curves are labeled in the same way 
as in Fig. \ref{fig:swave}.}
\end{figure}
The $J/\psi$ and $\Upsilon$ wave functions have the
structure indicated in Eq. (\ref{Upsilon}). The $s$-wave
wave function is denoted by $S$, and the $d$-wave wave
function by $D$. Both are functions of the relative momentum
$\vec k$ of the two effective quarks only through its
length, $k = |\vec k \,|$. This is a consequence of that the
wave functions $S$ and $D$ depend on the invariant mass
squared of the two quarks, ${\cal M}^2 = 4(m^2 + k^2)$. The
relative momentum $\vec k$ shares many properties with the
momentum-space variable typically introduced in
non-relativistic potential models, but one has to remember
that the variable $\vec k$ appears in QCD according to the
rules of LF dynamics.

The wave functions $S$ and $D$ are shown in Figs. \ref{fig:swave}
and \ref{fig:dwave} in four versions, two for the ground state
of $\Upsilon$ and two for $J/\psi$. Two versions per meson 
are obtained using the two choices of parameters $\alpha$ and 
$m$ that are given in Tables \ref{tab:bottom} and \ref{tab:charm}. 
Numerical values of the wave functions can be read from the 
tables given in Appendix \ref{wf}.

It is visible that the $d$-wave component is much larger 
in size in the charm case than in the bottom case, although
the $s$-wave components are similar in both cases at small 
relative momenta. This result can be attributed to much 
more relativistic relative motion of quarks in $J/\psi$ 
than in $\Upsilon$. Relativistic motion leads to enhancement 
of spin-dependent interactions that mix the $d$-wave component 
with the $s$-wave component.

A comment is in order regarding decay widths of the mesons.
In the leading approximation, the leptonic decay amplitudes 
are proportional to the $s$-wave wave functions at the origin 
in position space (integrals of the wave functions in momentum 
space). There is little doubt that the leptonic decay widths 
in the Hamiltonian approach to QCD will be qualitatively 
similar to the widths obtainable in potential models. On the 
other hand, inclusion of the term $\alpha^2 R$ in the effective 
potential and corresponding self-interactions in $H_{Q \bar 
Q \lambda}$, in a calculation similar to the simplest version 
discussed here, and further evaluation of the gluonic components 
in the effective-particle Fock-space basis in the eigenvalue 
problem for $H_\lambda$, may provide quantitative insight into 
effects not accessible in potential models. These effects may 
influence the leptonic decay rates and need to be evaluated 
in order to precisely compare the theory with data including 
the decay rates. The point is that such studies appear feasible 
using the boost-invariant Hamiltonian approach to QCD.
%
\section{ Conclusion }
\label{c}
The simplest approximate computation of masses of heavy quarkonia
in QCD with one flavor of quarks, still using the gluon mass-gap
ansatz to finesse a simple picture in the $|Q_\lambda \bar
Q_\lambda \rangle$ sector, suggests that the boost-invariant
Hamiltonian approach offers a feasible path to extended
studies of quark and gluon dynamics in the light-front Fock
space. One can use creation and annihilation operators for
effective particles in order to explicitly construct the states
of heavy quarkonia. The approach produces an approximate
constituent picture that is relativistic and usable for
description of fast-moving mesons. Masses of the mesons are
reproduced in the simplest version reasonably well for reasonable
values of the coupling constant, $\alpha$, and quark mass, $m$,
using a small set of basis states in the effective eigenvalue problem.
In the simplest version, the short-distance high-energy effects
and large-distance gluon dynamics are not fully described.
Therefore, it is not surprising that the simplest approach can
be used to reproduce only a small set of meson masses that lie 
in the middle of the spectra of small effective Hamiltonians 
called ``windows,'' see Section \ref{la}. But there is also no 
immediate reason found to question that the LF Hamiltonian 
approach to QCD may provide an interesting alternative to 
other approaches.

The Hamiltonian calculus produces the boost-invariant wave
functions that describe heavy quarkonia in terms of their
virtual effective-particle components in the LF Fock space.
In principle, these wave functions not only provide a
relativistic quantum image of a single hadron, but they also
can be used in description of decays, production, and
scattering of the quarkonia using QCD. Although the cases
discussed here concern only $b \bar b$ and $c \bar c$
systems, the extension to the case of unequal masses, such
as $b \bar c$ or $c \bar b$ mesons, requires only that
instead of the relative momentum variable $\vec k$ used here 
(see Appendix \ref{RGPEPscaling}) one uses the momentum 
variable defined in analogous way by the relations
\begin{eqnarray}
k ^\perp & = & \kappa^\perp \, , \\
\sqrt{ m_1^2 + \vec k ^{\, 2} } & + & \sqrt{ m_2^2 + \vec k ^{\, 2} } \nonumber \\
& = & 
\sqrt{ {m_1^2 + \kappa^{\perp \, 2} \over x_1 } 
+ 
{m_2^2 + \kappa^{\perp \, 2} \over x_2 } }\, .
\end{eqnarray}
Associated momentum-space techniques to handle two- and
three-particle systems with different masses in the context
of studies of the bound-state structure or decay are
sufficiently advanced in the LF approach to handle states
that contain quarks and gluons \cite{decay}. It is also
known that the gluon mass ansatz technique works reasonably
well in the case of gluonium \cite{TMaslowskiPhD}. Thus, it 
seems plausible that the case of different quark masses may 
be treated with explicit inclusion of the quark-antiquark-gluon 
sector. Knowing the corresponding wave functions, one can 
attempt to describe a host of new exclusive or semi-exclusive
processes that involve heavy quarkonia in arbitrary motion.

The formalism of LF dynamics in quantum field theory
involves a choice of an axis in space, especially in gauge
theories such as QCD, where one has to make a choice of
gauge depending on that axis. Therefore, the rotational
symmetry of the theory is not explicit in the LF Hamiltonian
formalism. Most of the expressions one encounters depend on
the distinguished axis. It is reassuring that the LF
Hamiltonian approach to heavy quarkonia produces in its
simplest version developed here explicit expressions for
bound-state spectra in which masses are exactly arranged in
multiplets corresponding to the total angular momentum
(meson spin) $J=0$, $J=1$, and $J=2$, and the wave functions
of the corresponding states are classifiable as waves $s$,
$p$, $d$, and $f$. Nevertheless, the complete expressions for
the wave functions contain additional relativistic factors
that are entirely outside the scope of non-relativistic
potential models, see Eqs. (\ref{state}), (\ref{wavefunction}), 
(\ref{wavefunctionCMF}), and Appendix \ref{additionalturn}.

The most attractive feature of the boost-invariant
Hamiltonian approach to heavy quarkonia, the one that makes
it an interesting candidate for a new expansion method in
solving QCD \cite{Wilson2004}, is that the renormalization
group procedure for effective particles can be
systematically studied order by order in expansion in powers
of the effective coupling constant $\alpha_\lambda$ with
$\lambda$ on the order of quark masses. Such expansion may
provide a reasonably converging sequence of approximations
if $\alpha_\lambda$ is much smaller than 1. This study shows
that $\alpha \sim 1/3$ is a reasonable candidate to
reproduce the masses of $b \bar b$ and $c \bar c$ mesons in
systematic calculations. Genuine 4th-order RGPEP studies
will further clarify if this hope is realistic.

On the other hand, careful readers have certainly observed
that the LF Hamiltonian dynamics with a harmonic oscillator
potential leads to the eigenvalues $M^2$ that are
proportional to the angular momentum in the relative motion
of quarks, like in the Regge trajectories. This is a
phenomenologically desired feature, although one cannot
trust the oscillator picture over large distances. But when
one considers highly excited states, their masses increasing
as dictated by the quadratic potential, the probability of
emission of effective gluons will be also increasing. A
string of gluons may be formed, with new potentials between
heavy effective gluons that require further investigation.
The Hamiltonian approach could thus lead to a quantum theory
of the gluon string and provide another reason for the same
Regge-like behavior of the spectrum, with a different slope
than implied by the two-quark approximation and with
validity extending to much larger distances than the size of
a typical hadron. In fact, for a firm chain of quantum
gluons to form a string, each pair of the neighboring gluons
must be held together stronger than by a linear potential,
and a quadratic potential satisfies this condition. The
pilot calculation described here suggests that the
oscillator frequencies are on the order of one inverse
fermi, and the oscillator potential term grows as the
relative distance squared in fermis with a coefficient given
by the quark mass. This means that the oscillator potential
is strong for the inter-quark distances larger than about a
fermi and the quantum theory of the gluon strings with a
similar potential between gluons may turn out to be useful
in phenomenology. Thus, the structure emerging in this pilot
study of the boost-invariant Hamiltonian approach to QCD has
a reasonable chance to grow toward a realistic physical
picture supported by a mathematically well-defined theory.
This is more than another reason to undertake the 4th-order
studies of the approach.

\vskip.2in
{\bf Acknowledgement}

This work was supported in part by the Grants No. 
KBN 1 P03B 117 26 and MNiSW BST-975/BW-1640.
%
\begin{appendix}
%
\section{ Terms in $H_\lambda$ }
\label{TermsinHlambda}
This appendix lists details of terms in Eq. (\ref{Hl}) for
$H_\lambda$. The kinetic energy term for effective quarks reads
\begin{equation}
\label{tql}
T_{q \lambda} = 
\sum_{\sigma c} \int [k] {k^{\perp \, 2} + m^2_\lambda \over k^+}
  \left[b^\dagger_{\lambda k \sigma c}b_{\lambda k \sigma c} + 
  d^\dagger_{\lambda k \sigma c }d_{\lambda k \sigma c} \right] \, ,
\end{equation}
and for certain $\lambda = \lambda_0$ one can set 
\begin{eqnarray}
\label{formofm0} 
m^2_{\lambda_0} & = & m^2 + 
{4 \over 3}  g^2 \, 
\int [x \kappa]  \, \tilde r^2_\delta(x)\, 
f^2_{\lambda_0} (m^2,{\cal M}^2)\, \nonumber \\ 
& \times & 
{   j^{\mu *}j^{\nu}\, 
    \left[ - g_{\mu \nu} \, + \, 
    { n_\mu n_\nu  \over p^{+ \, 2} } \, { {\cal M}^2  -  m^2 \over x }  
   \right]
 \over {\cal M}^2 - m^2} \, ,
\end{eqnarray}
with ${\cal M}^2 = \kappa^{\perp \, 2}/x + (\kappa^{\perp \, 2}
+ m^2)/(1-x)$. The integration measure $[x\kappa]$ stands for 
$dx \, d^2 \kappa^\perp /[ 16\pi^3 x(1-x)]$. That the effective 
mass does not depend on the particle motion is a unique property 
of the RGPEP in LF dynamics. $\tilde r_\delta (x)$ denotes the 
small-$x$ regularization factors $r_\delta (x) \, r_\delta (1-x)$, 
where $r_\delta(x) = x^\delta \, \theta(x)$. The gluon kinetic 
energy term reads,
\begin{equation}
\label{tgl}
T_{g \lambda} = \sum_{\sigma c} \int [k] {k^{\perp \, 2} + \mu^2_\lambda 
\over k^+} \, a^\dagger_{\lambda k \sigma c}a_{\lambda k \sigma c} \, ,
\end{equation}
but an explicit expression for $\mu^2_\lambda$ \cite{RGPEP, glambda} is not 
needed in this work.

The emission and absorption term, $Y_\lambda = f_\lambda Y_{qg \lambda}$, is 
\begin{eqnarray}
\label{yl}
Y_\lambda & = \, g \sum_{123} \, \int[123]  
\,r_\delta(x_{1/3})\,r_\delta(x_{2/3})
\, f_\lambda({\cal M}^2_{12},m^2) \nonumber \\
& \times \left[j_{23} \, b^\dagger_{\lambda 2} 
  a^\dagger_{\lambda 1} b_{\lambda 3} +
  \bar j_{23} \, d^\dagger_{\lambda 2} 
  a^\dagger_{\lambda 1} d_{\lambda 3} + 
h.c. \right] \, ,
\end{eqnarray} 
where $j_{23} = \tilde \delta \, t^1_{23} \, 
g_{\mu \nu} \, j^\mu_{23} \,  \varepsilon^{\nu \, *}_1$,
$\bar j_{23} = \tilde \delta \, t^1_{32} \, 
g_{\mu \nu} \, \bar j^\mu_{32} \, \varepsilon^{\nu \, *}_1$,
$\tilde \delta$ denotes the $\delta$-function of 
three-momentum conservation times $16 \pi^3$, $t^a$ 
with $a = 1, ..., 8$ denote $3 \times 3$ matrices of 
generators of color $SU(3)$ gauge transformations 
for quarks, $\varepsilon$ is the gluon polarization 
four-vector, $j^\mu_{23} = \bar u_2 \gamma^\mu u_3$ 
and $\bar j^\mu_{32} = \bar v_3 \gamma^\mu v_2$. 

\begin{figure}[ht]
\includegraphics[scale=.9]{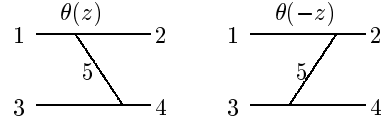}
\caption{\label{fig:oge} Momentum labels in potential terms.}
\end{figure}

The potential term, $V_\lambda = f_\lambda V_{q \bar q \lambda}$, is 
\begin{equation}
\label{vl}
V_\lambda = -g^2 \,\sum_{1234}\int[1234] 
\, \tilde\delta \, t^a_{12} t^a_{43} V_\lambda(13,24) 
\, b_1^\dagger d_3^\dagger d_4 b_2 \, ,
\end{equation}
where (see Ref. \cite{ho})
\begin{eqnarray}
 V_\lambda(13,24) & = & 
 { d_{\mu \nu}(k_5) \over k^+_5} 
 j^\mu_{12} \bar j^\nu_{43}  \,
 f_\lambda({\cal M}^2_{13}, {\cal M}^2_{24})     \\
&\times & \left[
  \theta(z)
  \tilde r_\delta (x_{5/1}) \tilde r_\delta (x_{5/4}) \,
  {\cal F}_{2 \lambda} (1, 253, 4) \right.          \nonumber \\
&  +    & \left.
  \theta(-z)
  \tilde r_\delta (x_{5/3}) \tilde r_\delta (x_{5/2})
  \,{\cal F}_{2 \lambda} (3, 154, 2) 
        \right] \, ,                                \nonumber 
\end{eqnarray}
and, for example, 
\begin{eqnarray}
\label{calFexample}
& & { \mathcal{F}_{2\lambda}(1,253,4) \over P^+ }
= 
{ x_1({\cal M}_{52}^2 - m^2) + x_4({\cal M}_{53}^2 - m^2) \over 
({\cal M}_{52}^2 - m^2)^2 + ({\cal M}_{53}^2 - m^2)^2 } \nonumber \\
&\times &
\left[  \exp{\left[ - 
{({\cal M}_{52}^2 - m^2)^2 + ({\cal M}_{53}^2 - m^2)^2
\over \lambda^4 } \right]}
      - 1 \right].
\end{eqnarray}
${\cal M}_{ij} = (k_i + k_j)^2$ is the invariant mass of 
particles $i$ and $j$.
Momenta are labeled according to Fig. \ref{fig:oge}.
$P^+$ denotes the sum of plus momenta of annihilated quarks. 
The sum over gluon polarizations, 
\begin{equation}
\label{dmunu}
d_{\mu \nu}(k_5) = - g_{\mu \nu} + 
{ n^\mu k_5^\nu + k_5^\mu n^\nu \over k_5^+ } \, ,
\end{equation}
involves momentum $k_5^{+, \perp} = \varepsilon(z) 
\, (k_1^{+, \perp} - k_2^{+, \perp})$ with  
$\varepsilon(z) = \theta(z) - \theta(-z)$, and 
$z = (k_1^+ - k_2^+)/(k_1^+ + k_3^+)$. $x_5 = |z|$,
and $k_5^- = k_5^{\perp \, 2}/ k_5^+$. The instantaneous 
interaction between effective quarks, $Z_\lambda = 
f_\lambda Z_{q\bar q \lambda}$, is 
\begin{eqnarray}
\label{zl}
Z_\lambda = -g^2 \sum_{1234}\int[1234] 
\tilde\delta t^a_{12} t^a_{43} Z_\lambda(13,24)
            \, b_{\lambda 1}^\dagger d_{\lambda 3}^\dagger 
               d_{\lambda 4} b_{\lambda 2} , \nonumber \\
\end{eqnarray}
where
\begin{eqnarray}
\label{zzl}
 Z_\lambda(13,24) & = & {1 \over k_5^{+ \, 2}} \, 
j^+_{12} \bar j^+_{34} \,
f_\lambda({\cal M}^2_{13}, {\cal M}^2_{24})  \nonumber \\
& \times & 
[  \theta(z)\tilde r_\delta(x_{5/1}) \tilde r_\delta(x_{5/4}) \nonumber \\
& + &
\theta(-z)\tilde r_\delta(x_{5/3}) \tilde r_\delta(x_{5/2})] \, . 
\end{eqnarray}
\section{ RGPEP scaling with $\alpha$ }
\label{RGPEPscaling}
This appendix includes contributions that
originate in the $g_{\mu \nu}$-parts of the
sums over gluon polarizations, which were not
explicitly described in Ref. \cite{ho}.

The analysis of scaling with $\alpha_0$ for
RGPEP factors in the window eigenvalue
equation for $H_{\lambda_0}$ with a mass
ansatz $\mu^2$ is based on the similarity
between the structure of the eigenvalue Eq.
(\ref{MainEq}) and the Schr\"odinger equation
with Coulomb potential in QED (the same kind
of the leading picture is also found in
Yukawa theory \cite{largep}). The subscripts
$\lambda_0$ and $0$ are often omitted to simplify
notation. 

When the relative momentum of electron and 
positron in positronium is written as
\begin{eqnarray}
\vec k & = & \alpha \, \mu \, \vec p \, , 
\end{eqnarray}
where $\mu$ is the reduced mass of the 
fermions, the Schr\"odinger equation for 
positronium, neglecting spin effects, 
\begin{eqnarray}
            {k^2 \over 2 \mu} \, \psi(\vec k)
       - \, \int { d^3 k' \over (2\pi)^3 } \,
             { 4 \pi \, \alpha \over (\vec k - \vec k \, ')^2 } \, 
             \psi(\vec k \, ') 
& = & E \psi(\vec k) \, ,
\end{eqnarray}
takes the form 
\begin{eqnarray}
p^2 \, \phi(\vec p)
       - 2 \, \int { d^3 p' \over (2\pi)^3 } \,
             { 4 \pi \over (\vec p - \vec p \, ')^2 } \, 
             \phi(\vec p \, ') 
& = & x \, \phi(\vec p) \, .
\end{eqnarray}
The eigenvalue is $E= x \, {1 \over 2} \, 
\mu \, \alpha^2$, and the ground-state has 
the eigenvalue $E = E_0$ with $x = x_0 = -1$ 
and wave function
\begin{eqnarray}
\phi_0(\vec p)
      & = & 
{ N_p \over (p^2 + 1)^2 } \, .
\end{eqnarray}
Higher states have $x= -1/n^2$ with natural 
$n$ greater than 1. 

In the QCD Schr\"odinger equation with $H_{Q \bar Q
\lambda}$, the self-interaction terms and the
potential kernel contain similar expressions.
The self-interaction terms are easy to
analyze if one knows how to analyze the structure
of $v_\lambda$ in the eigenvalue Eq.
(\ref{MainEq}). At certain $\lambda =
\lambda_0$, with $\lambda_0$ parameterized
according to Eq. (\ref{lambda0}), one has
$v_{\lambda_0}(13,24) = v_0(13,24)$ and 
\begin{eqnarray}
\label{bracket}
v_0  & = & - A \,g_{\mu \nu} \,
j_{12}^\mu \bar j_{43}^\nu 
+
B \, { j_{12}^+ \bar j_{43}^+ \over P^{+ \, 2} } \, , \\
\label{defineA}
A 
& = &  { 1\over |z|} \, 
\left[ f_0(13,24)\, {V\over P^+}   \,
+{1 \over 2} \, w_0(13,24) \right] \, , \\
\label{defineB}
B
& = &
{ 1 \over z^2 } \, 
{d \over |z|} \,\left[ f_0(13,24)\, {V \over P^+} \,
+{1 \over 2} \, w_0(13,24)\right] \nonumber \\ 
& + &
{ 1 \over z^2 } \, f_0 (13,24)\, Z 
\, ,
\end{eqnarray}
where
\begin{eqnarray}
V & = &  
  \theta(z)
  \tilde r_\delta (x_{5/1}) \tilde r_\delta (x_{5/4}) \,
  {\cal F}_{2 \lambda} (1, 253, 4)        \nonumber \\
&  +    & 
  \theta(-z)
  \tilde r_\delta (x_{5/3}) \tilde r_\delta (x_{5/2})
  \,{\cal F}_{2 \lambda} (3, 154, 2) \, ,      \\
Z & = & 
 \theta(z)\tilde r_\delta(x_{5/1}) \tilde r_\delta(x_{5/4}) 
\nonumber \\
& + &
\theta(-z)\tilde r_\delta(x_{5/3}) \tilde r_\delta(x_{5/2})  \, , \\ 
\label{w0appendix}
{ w_0 \over |z| }
& = & 
 { \theta(z)\,
  \tilde r_\delta (x_{5/1}) \tilde r_\delta (x_{5/4}) \,
  f_{52} f_{53}  
\over 
|z| (m^2 - {\cal M}_{52}^2)/x_1 - \mu^2(2,5,3)  }  \nonumber \\
& + &  
 { \theta(-z)\,
  \tilde r_\delta (x_{5/3}) \tilde r_\delta (x_{5/2}) \,
  f_{54} f_{51}
\over
|z| (m^2 - {\cal M}_{54}^2)/x_3 - \mu^2(1,5,4)   }  \nonumber \\
& + &  
 { \theta(z)\,
  \tilde r_\delta (x_{5/1}) \tilde r_\delta (x_{5/4}) \,
 f_{52} f_{53}  
\over 
|z| (m^2 - {\cal M}_{53}^2)/x_4 - \mu^2(2,5,3)   }  \nonumber \\
& + &  
 { \theta(-z)\,
  \tilde r_\delta (x_{5/3}) \tilde r_\delta (x_{5/2}) \,
  f_{54} f_{51}
\over
|z| (m^2 - {\cal M}_{51}^2)/x_2 - \mu^2(1,5,4)   }  \, , \\
{d \over |z|} 
& = &
\theta(z)
  \left( { {\cal M}_{52}^2 - m^2 \over 2 x_1} 
                + { {\cal M}_{53}^2 - m^2 \over 2 x_4} \right) \nonumber \\
& + &
\theta(-z) 
  \left( { {\cal M}_{51}^2 - m^2 \over 2 x_2}
                + { {\cal M}_{54}^2 - m^2 \over 2 x_3} \right) \, ,
\end{eqnarray}
and $f_{ij}$ denotes $f_{\lambda_0}[m^2,(k_i+k_j)^2]$.

In order to describe the structure of $v_0$
for relative quark momenta comparable with
the strong Bohr momentum, introduced in Eq.
(\ref{kB}), it is convenient to write
expressions for ${\cal F} (1,253,4)/k_5^+$
and ${\cal F}(3,154,2)/k_5^+$ that contribute
to $V$ using identities
\begin{eqnarray}
\mathcal{M}_{253}^{2} & = & 
\frac{\mathcal{M}_{52}^{2}-m^{2}}{x_{1}}+\mathcal{M}_{13}^{2}
=\frac{\mathcal{M}_{53}^{2}-m^{2}}{x_{4}}+\mathcal{M}_{24}^{2}\,,
\nonumber \\
& & \\
\mathcal{M}_{154}^{2} & = & 
\frac{\mathcal{M}_{54}^{2}-m^{2}}{x_{3}}+\mathcal{M}_{13}^{2}
=\frac{\mathcal{M}_{51}^{2}-m^{2}}{x_{2}}+\mathcal{M}_{24}^{2}\,.
\nonumber \\
&& 
\end{eqnarray}
One starts with expressions, see Eq. (\ref{calFexample}),
\begin{eqnarray}
{{\cal F}_2(1,253,4) \over k_5^+ } & = & 
\frac{1}{\left|z\right|}\frac{x_{1}\left(\mathcal{M}_{52}^{2}-m^{2}\right)
+x_{4}\left(\mathcal{M}_{53}^{2}-m^{2}\right)}{\left(\mathcal{M}_{52}^{2}-m^{2}\right)^{2}
+\left(\mathcal{M}_{53}^{2}-m^{2}\right)^{2}} \nonumber \\
& \times & \left(ff-1\right)\,,\\
{{\cal F}_2(3,154,2) \over k_5^+ } & = & 
\frac{1}{\left|z\right|}\frac{x_{2}\left(\mathcal{M}_{51}^{2}-m^{2}\right)
+x_{3}\left(\mathcal{M}_{54}^{2}-m^{2}\right)}{\left(\mathcal{M}_{51}^{2}-m^{2}\right)^{2}
+\left(\mathcal{M}_{54}^{2}-m^{2}\right)^{2}} \nonumber \\ 
& \times & \left(ff-1\right)\, ,
\end{eqnarray}
where
\begin{eqnarray}
\mathcal{M}_{51}^{2}-m^{2} & = & \mathcal{D}_{1}\,,\\
\mathcal{M}_{52}^{2}-m^{2} & = & \frac{x_{1}}{x_{2}}\mathcal{D}_{1}\,,\\
\mathcal{M}_{53}^{2}-m^{2} & = & \mathcal{D}_{3}\,,\\
\mathcal{M}_{54}^{2}-m^{2} & = & \frac{x_{3}}{x_{4}}\mathcal{D}_{3}\,,
\end{eqnarray}
and
\begin{eqnarray}
{\cal D}_1 & = & {x_1 \over |z|}
\left[\left(q^\perp - {z \over x_1} k_{13}^\perp \right)^2
+ m^2 {z^2 \over x_1^2} \right] \, , \\
{\cal D}_3 & = & {x_3 \over |z|}
\left[\left(q^\perp + {z \over x_3} k_{13}^\perp \right)^2
+m^2  {z^2 \over x_3^2} \right] \, .
\end{eqnarray}
The definitions include
\begin{eqnarray}
z   & = & x_1 - x_2 \, , 
\end{eqnarray}
and it is helpful to use three-dimensional CMF relative momentum 
variables $\vec k_{13}$ and $\vec k_{24}$, and $\vec q = 
\vec k_{13} - \vec k_{24}$. So, for $ij=13$ and $24$,
\begin{eqnarray}
x_i  & = & {1 \over 2} + {k_{ij}^3 \over 2 \sqrt{m^2 + \vec k_{ij}^{\,2}} } \, , 
\end{eqnarray}
and
\begin{eqnarray}
z & = &  {k_{13}^3 \over {\cal M}_{13} } - {k_{24}^3 \over {\cal M}_{24} } \, .
\end{eqnarray}

The first step is to establish that the
potential does not generate any small-$x$
singularities in its fully relativistic form
\cite{ho}. The next step is to analyze
scaling with $\alpha$. The key to scaling
with $\alpha$ for given quark mass $m$ is the
substitution 
\begin{eqnarray}
\label{scalingsubstitution}
\vec k_{ij} & = & k_B \,\, \vec p_{ij} \, ,
\end{eqnarray}
where $k_B$ is the strong Bohr momentum of
Eq. (\ref{kB}). The dimensionless variables
$\vec p_{ij}$, with $ij$ = 13 or 24, are
typically on the order of 1 in both the
purely Coulombic case of QED and in the QCD 
case that includes the harmonic oscillator 
potential studied here. A dimensionless momentum
transfer $\vec p$ is defined by 
\begin{eqnarray}
\label{scalingp}
\vec q & = & k_B \,\, \vec p \, ,
\end{eqnarray} 
so that $\vec p = \vec p_{13} - \vec p_{24}$.
Factors $ff$ limit $|\vec p \, |$ to values
order $(4 \alpha/3)^{2 \epsilon}
\lambda_p^2$, and the additional damping due
to $\epsilon > 0$ provides a possibility to
formally separate the dominant terms in the
limit $\alpha \rightarrow 0$ because the
Coulomb eigenvalue problem is dominated by
the dimensionless momenta $p_{ij}$ on the
order of 1. The outer-most factor $f_0$ in
the potential terms limits changes of momenta
$p_{ij}$ from above by
$\alpha^{\epsilon-0.5}$ and this $f_0$
becomes irrelevant for very small $\alpha$,
leaving the Coulomb interaction and the
harmonic oscillator term that provide the
leading approximation.

Observe that 
\begin{eqnarray}
{ {\cal F}_2(1,253,4) \over k_5^+ (ff-1)} 
 & = & { ( {\cal M}_{253}^2 - C_{253}) \, |z|^{-1} \over {\cal M}_{253}^4 - 2 {\cal M}_{253}^2 C_{253} + D_{253} } \, , \\
{ {\cal F}_2(3,154,2) \over k_5^+ (ff-1)} 
 & = & { ( {\cal M}_{154}^2 - C_{154}) \, |z|^{-1} \over {\cal M}_{154}^4 - 2 {\cal M}_{154}^2 C_{154} + D_{154} } \, ,
\end{eqnarray}
where 
 \begin{eqnarray}
C_{253} & = &  { x_1^2 \mathcal{M}_{13}^2 + x_4^2 \mathcal{M}_{24}^2  \over  x_1^2 + x_4^2 }  \, , \\
D_{253} & = &  { x_1^2 \mathcal{M}_{13}^4 + x_4^2 \mathcal{M}_{24}^4  \over  x_1^2 + x_4^2 }  \, , \\
C_{154} & = &  { x_3^2 \mathcal{M}_{13}^2 + x_2^2 \mathcal{M}_{24}^2  \over  x_3^2 + x_2^2 }  \, , \\
D_{154} & = &  { x_3^2 \mathcal{M}_{13}^4 + x_2^2 \mathcal{M}_{24}^4  \over  x_3^2 + x_2^2 } \, .
\end{eqnarray}
Using Eq. (\ref{scalingsubstitution}), and introducing 
two three-vectors, 
\begin{eqnarray}
\label{xivector}
\vec \xi & = & { \vec{k}_{13} + \vec{k}_{24} \over m } = O(\alpha) \, , \\
\label{evector}
\vec \iota   & = & \vec q / q \, ,
\end{eqnarray}
one obtains
\begin{eqnarray}
{\cal M}_{253}^2 - C_{253}
& = & 
{\cal M}_{154}^2 - C_{154} + { O(\alpha^5) \over |z| } \\
& = &  
{ q^2 \over |z|} \, 
\left( 1 - \iota_z \xi_z \, \vec \iota \vec \xi \, \right) + { O(\alpha^5) \over |z|} \, ,  \\
D_{253} - C_{253}^2
& = & 
D_{154} - C_{154}^2 + { O(\alpha^7) \over z^2 }  \\
& = & 
{q^4  \over z^2 } \, \iota_z^2 \, \left( \vec \iota \vec \xi \, \right)^2 + { O( \alpha^7 ) \over z^2 } \, .
\end{eqnarray}
Thus,
\begin{eqnarray}
{ \mathcal{F}_2(1,253,4) \over k_5^+(ff-1) }  
 & = & 
{ \mathcal{F}_2(3,154,2) \over k_5^+(ff-1) } + O(\alpha)  \\
= {1 \over q^2 } & + & 
{ \iota_z \xi_z \, \vec \iota \vec \xi -  \iota_z^2 \, ( \vec \iota \vec \xi )^2 \over q^2 } 
+ O(\alpha) 
\, .
\end{eqnarray}
The effective-gluon exchange term $w_0$ in Eq. (\ref{w0appendix}), is
\begin{eqnarray}
{ w_0 \over |z| }
& = &
- { 2 \theta(z)\,
  \tilde r_\delta (x_{5/1}) \tilde r_\delta (x_{5/4}) \, f_{52} f_{53}  
\over 
q^2 + \mu^2(2,5,3) + O(\alpha^{3 + 4 \epsilon}) }  \nonumber \\
& - & 
{ 2 \theta(-z)\,
  \tilde r_\delta (x_{5/3}) \tilde r_\delta (x_{5/2}) \, f_{54} f_{51}
\over
q^2 + \mu^2(1,5,4) + O(\alpha^{3 + 4 \epsilon}) }  \, ,
\end{eqnarray}
and the intermediate gluon spin sum contributes
\begin{eqnarray}
d  
& = & 
q^2 + O(\alpha^6) \, .
\end{eqnarray}

In summary, the factors $A$ and $B$, 
defined in Eqs. (\ref{defineA}) and 
(\ref{defineB}), scale as 
\begin{eqnarray}
\label{Aappendix}
A & \simeq & - f_0 \, { 1 \over q^2 } \, 
\left[ 1-{ff \over f_0} \, 
\left(f_0- {q^2 \over q^2 + \mu^2 }\right) + c \right]
 \, , \\
\label{Bappendix}
B & \simeq & - f_0 \, {4m^2 \over q_z^2} \,
\left[ - {ff \over f_0} \,  
\left(f_0- {q^2 \over q^2 + \mu^2 }\right) + c \right] \, , \\
c 
& = &
\vec e_z \vec \iota  \quad \vec \iota \vec \xi  \,\,
\left( \vec e_z \vec \xi  - \vec e_z \vec \iota \,\,\, \vec \iota \vec \xi \, \right) 
\, + O(\alpha^3)
\, ,
\end{eqnarray}
and $c \sim \alpha^2$ because $|\vec \xi \,|
= O(\alpha)$. These scaling results are valid
even if the mass ansatz $\mu^2$ is of the
order of $\alpha$ instead of 1.

Scaling analysis of the self-interaction
terms begins with the RGPEP expression for
renormalized effective quark mass terms in
the eigenvalue Eq. (\ref{MainEq}),
\begin{eqnarray}
{\delta m^2_i \over x_i}
& = &
{4 g^2 \over 3 x_i} \int [y\rho]\,
f^2(m^2,\mathcal{M}^2) \,
{ 2 \over  1-y} \nonumber \\
& \times & 
\left\{ m^2 y^2 + \left[ 1 + (1 - y)^2  \right] \left( {\rho^\perp  \over y} \right)^ 2 \right\} 
\nonumber \\ 
& \times & 
\left[ {1 \over \mathcal{M}^2 - m^2 } - {1 \over \mathcal{M}_i^2 - m^2 } \right] \, .
\end{eqnarray}
where 
\begin{eqnarray}
{\cal M}_i^2 - m^2 
& = & 
{\cal M}^2 - m^2 + {\mu^2 \over y}  \, , \\ 
\mathcal{M}^2 - m^2 
& = & 
{\rho^{\perp \, 2} + y^2 m^2 \over y (1-y) } \, ,
\end{eqnarray}
and $\mu^2$ is the mass ansatz for the 
effective gluon accompanying the quarks 
$i'$ and $j$. One can introduce the 
variable 
\begin{eqnarray}
q^\perp = \rho^\perp \, , \quad q_z = y m \, ,
\end{eqnarray}
and observe that when $\lambda_0 \sim
\alpha^{0.5 + \epsilon} m$, the magnitude of
$q$ is limited by the RGPEP form factor $f^2$
to the range between 0 and
$\alpha^{1+2\epsilon} m$. Corrections order
$y$ cancel out or multiply terms order $q^2$.
If $\mu^2$ is of order 1, deviations from
$y=0$ introduce corrections order $\alpha^{4
+ 6\epsilon}$ and can be formally neglected
when one keeps only terms order $\alpha^2$ and
$\alpha^4$. Note that the terms order $\alpha^3$ 
cancel out completely, and even for $\mu^2$ order
$\alpha$ the first correction due to $y \neq
1$ is order $\alpha^{4+6\epsilon}$. Then,
since the integrand is symmetric in $t =
\cos{\theta}$, where $q_z = q t$, and since
$x_1 x_2$ differs from 1/4 in order
$\alpha^2$, the leading contribution from the
self-interactions is
\begin{eqnarray}
\label{selfinteractionappendix}
{\delta m^2_1 \over x_1}
+
{\delta m^2_3 \over x_3}
& = &
8 m  {4 \over 3} {g^2 \over 16 \pi^3} 
\int d^3 q \nonumber \\
& \times &
ff
\left( {1 \over  q_z^2 } - {1 \over  q^2 } \right) \,
{\mu^2  \over \mu^2 + q^2 }
 \, .
\end{eqnarray}
\section{ Spinors }
\label{additionalturn}
The $4 \times 4$ matrix wave function $\Psi_{ij}$ in Eq.
(\ref{wavefunction}), is written using spinors $u_{k,s}$ 
and $v_{k,s}$ that are obtained by applying matrix $B(k,m)$ 
from Eq. (\ref{Bkm}) to spinors at rest, $u_{0,s}$ and
$v_{0,s}$, which are defined as
\begin{eqnarray}
u_{0,s} & = & \sqrt{2m} \left[
\begin{array}{c}
\chi_s\\
0\end{array}
\right] \, , \\
v_{0,s} & = & 2s \sqrt{2m} \left[
\begin{array}{c}
0\\
\chi_{\bar s}\end{array}
\right] \, . 
\end{eqnarray}
$\chi_s$ denotes standard two-component
spinors, with upper component equal $1/2+s$
and lower $1/2-s$, and $\bar s$ means $-s$.
Instead of $s = \pm 1/2$, $\sigma = 2s$ with 
values $\pm 1$ is often used below.
The above spinors correspond to fermions at
rest in the frame of reference in which one
carries out the calculation for the bound
state with momentum components $P^+$ and
$P^\perp$. The discussion below is simplified
to the case $P^\perp = 0$ since the
transverse motion of the bound state does not
introduce any change in the final formulas. 

The wave function $\Psi(\vec k_{ij}, s_i,
s_j)$ in Eq. (\ref{wavefunctionCMF}) is
defined using the matrix $\Psi_{CMFij}(\vec
k_{ij})$ and spinors $u_{\vec k_{ij}, s_i}$
and $v_{-\vec k_{ij}, s_j}$ that describe the
fermions in the CMF of the quarks $i$ and
$j$. These are obtained by standard matrices
for boosts along $\vec k_{ij}$ instead of the
LF boosts. Namely, 
\begin{eqnarray}
\label{meloshu}
u_{\vec k, \sigma} & = & 
L(\vec{k}) \, \sqrt{2m} \left[
\begin{array}{c}
\tilde{\chi}_{\vec k \, \sigma}\\
0\end{array}
\right] \, , \\
\label{meloshv}
v_{\vec k, \sigma} & = & 
L(\vec{k}) \, \sigma \sqrt{2m} \left[
\begin{array}{c}
0\\
\tilde{\chi}_{\vec k \, \bar{\sigma}}\end{array}
\right] \, , \\
L(\vec{k}) & = &
{ 1 \over \sqrt{2m(E_k + m)} } \, \left[ \not \! k + \beta m \right] \beta \, .
\end{eqnarray}
The two component spinors at rest are turned away from 
the $z$-axis using a $2 \times 2$ matrix $\zeta(\vec k)$:
\begin{eqnarray}
\label{chitilde}
\tilde{\chi}_{\vec k \, \sigma} & = &
\zeta(\vec k) \, \chi_\sigma  \, , \\
\zeta(\vec k) & = &
\sqrt{{k^+ \over m}}\,\frac{\sqrt{2m(E_k+m)}}
{\left(k^+ + m\right)^2 + k^{\perp\, 2}} 
 \left(k^+ + m + \sigma^\perp k^\perp \sigma^3 \right). \nonumber \\
& & 
\end{eqnarray}
The matrix $\zeta$ introduces the spinor basis in which the
wave function $\Psi_{CMFij}(\vec k_{ij})$ satisfies a
rotationally symmetric eigenvalue equation in the leading
approximation. The matrix $\zeta$ has been used before by
Melosh \cite{Melosh1} as a candidate for description of how
the constituent quarks are related to current quarks, and by
Brisudova and Perry~\cite{Melosh2} in LF eigenvalue problems. 
Here, the constituent quarks are dynamically related to current
quarks using RGPEP and $\zeta$ plays only the kinematical role
of choosing a basis for spinors. The matrix $\zeta$ is needed 
because there is a change of frame of reference involved in 
expressing LF spinors in the frame of reference in which
the whole quarkonium has momentum $P^+$ and $P^\perp = 0$ in 
terms of the spinors in the CMF of the pair of quarks. For 
example,
$u_{k_1,s_1} = {\cal L}_{13}
u_{\vec{k}_{13},s_1}$, where 
\begin{eqnarray}
{\cal L}_{ij} & = & 
\Lambda^+  \sqrt{ P^+ \over {\cal M}_{ij} }
+
\Lambda^-  \sqrt{ {\cal M}_{ij} \over P^+} \, .
\end{eqnarray}
When one uses the slightly rotated basis for the two-component 
spinors in the CMF of fermions, as indicated in Eqs. (\ref{meloshu}) 
and (\ref{meloshv}), and than calculates the spinors in the frame of
reference where the bound state eigenvalue $P^-$ is calculated, one 
obtains spinors that are used in Eq. (\ref{state}). E.g.,
\begin{eqnarray}
u_i
& = &
{\cal L}_{ij} \, B(k_{ij},m) \, \zeta^{-1}(\vec{k}_{ij}) 
L^{-1}(\vec{k}_{ij}) u_{\vec k_{13}, s_1} \, .
\end{eqnarray}
The matrix $\zeta$ is defined to render
\begin{eqnarray}
B(k_{13},m) \, \zeta^{-1}(\vec{k}_{13}) L^{-1}_{BD}(\vec{k}_{13}) 
& = & 1 \, ,
\end{eqnarray}
and $u_1 = {\cal L}_{13} \, 
u_{\vec k_{13}, s_1}$. Similarly,
$u_2 = {\cal L}_{24} \, u_{\vec k_{24}, s_2} $,  
$v_3 = {\cal L}_{13} \, v_{-\vec k_{13}, s_3}$, and 
$v_4 = {\cal L}_{24} \, v_{-\vec k_{24}, s_4}$.
\section{ Breit-Fermi terms }
\label{BFterms}
In the leading approximation for small $\alpha$, the 
potential kernel given in Eq. (\ref{v0}) is 
\begin{eqnarray}
\label{v0approximate}
v_0
& = &  
f \, {1 \over q^2 } \, \,g_{\mu \nu} \, j_{12}^\mu \bar j_{43}^\nu  \nonumber \\
& + &  
ff \, {\mu^2 \over q^2 + \mu^2 } \,
\left[ 
{ 4m^2 \,  j_{12}^+ \bar j_{43}^+ \over P^{+ \, 2} q_z^2 }
-
{ g_{\mu \nu} \, j_{12}^\mu \bar j_{43}^\nu \over q^2 } \, 
\right] \, .
\end{eqnarray}
The Breit-Fermi terms in this article
originate from the first term. Note that the
first term contains only one form factor and
this form factor limits the change of the
square of the dimensionless momentum $p_{ij}$
in the scaling analysis by a number of the
order of $\alpha^ {2\epsilon-1}$, which is
much larger than 1 when $0 < \epsilon < {1
\over 2}$, see Appendix \ref{RGPEPscaling}.
In contrast, the second term contains two
form factors and in the scaling analysis these 
form factors limit the dimensionless momentum 
transfer $p = |\vec p \, |$ between quarks by 
a small number on the order of $(|p_z|/p) 
\lambda_p^2 (4\alpha/3)^{2 \epsilon}$, see 
Section \ref{eqstructure} and Appendix
\ref{RGPEPscaling}. This difference between
the form factors implies, that in the first
term the dominant momentum scale is of order
1, originating from the Coulomb potential,
while in the second term the allowed momentum
exchange $p$ is in principle infinitesimal
as long as $\epsilon > 0$. Thus, the second 
term would be negligible in the scaling analysis 
if it did not contain the diverging factor 
$q_z^{-2}$. This divergence is regulated by 
the falloff of the ansatz $\mu^2$ as a function 
of $q_z$ when $q_z$ tends to zero, as required 
by the condition that the ansatz does not 
produce a small-$x$ divergence \cite{ho}. 

In the leading approximation, the current
factors are diagonal in spin: $j_{12}^\mu
\bar j_{43 \, \mu}$ equals $4m^2$, and $j_{12}^+
\bar j_{43}^+$ always equals $4 P^{+ 2}$
times $\sqrt{x_1 x_2 x_3 x_4}$. The square
root reduces to 1/4 since all the $x$s differ
from 1/2 only by terms of order $\alpha$ or
smaller. In this case, the second term in Eq.
(\ref{v0approximate}) is the same as the
integrand in Eq. (\ref{selfinteractionappendix})
for self-interactions. The self-interaction  
and the second term in Eq. (\ref{v0approximate}) 
produce together the harmonic oscillator potential 
in Eq. (\ref{MainEqphi}) with the dimensionless 
spring constant given in Eq. (\ref{kp}) \cite{ho}. 

Beyond the leading order, one has to analyze
the factor $j_{12}^\mu \bar j_{43 \, \mu}$.
It can be re-written in a matrix notation of
Eq. (\ref{BFinV}). The $BF$ terms in the 
potential kernel $\cal V$ act on the $2 \times 
2$ matrix wave function $\phi$ which is defined 
as follows. Using results from Appendix
\ref{additionalturn}, the sum over quark
spins in Eq. (\ref{MainEq}) can be written as
\begin{eqnarray}
\label{spin1}
& & \sum_{s_2 s_4} 
j_{12}^\mu \bar j_{43 \, \mu} \, 
\Psi_{s_2 s_4}(\vec k_{24})  \, \, =  \\
& & 
\bar u_{\vec k_{13},s_1}
\gamma^0  {\cal L}_{13}^\dagger \gamma^0 \, 
\gamma^\mu \, 
{\cal L}_{24} \, 
K_{24} \, 
\gamma^0 {\cal L}_{24}^\dagger \gamma^0 \, 
\gamma_\mu \, 
{\cal L}_{13} \, v_{-\vec{k}_{13},s_{3}}\, ,  \nonumber \\
& & 
\label{K1}
K_{ij} = \left(\not\! k_{ij}+m\right) \,
    \Psi_{CMFij}(\vec{k}_{ij}) \,
    \left(\not\! \bar k_{ij}-m\right) \, ,
\end{eqnarray}
where $k = (k^0, \vec k)$, 
      $\bar k = (k^0, - \vec k)$, 
and   $k^0 = \sqrt{ m^2 + \vec k^{\, 2}}$.
This $4 \times 4$ matrix notation can be
replaced by a $2 \times 2$ matrix notation 
using
\begin{eqnarray}
\label{wavefunctionCMFphi}
\Psi_{s_i s_j}(\vec k_{ij}) & = & 
\bar u_{\vec k_{ij}, s_i} \,
\Psi_{CMFij}(\vec k_{ij}) \, v_{-\vec k_{ij},
s_j}  \nonumber \\
& \equiv &
\label{phi1}  
\sqrt[4]{ 1 + \vec k_{ij}^{\,2}/m^2}  \,\,
\sigma_j \tilde \chi_i^\dagger 
\phi (\vec p_{ij}) \tilde \chi_{\bar j} \, ,
\end{eqnarray}
where 
\begin{eqnarray}
\label{phi2}
\phi (\vec p_{ij}) & = & a + \vec b \cdot \vec{\sigma} 
\end{eqnarray}
is the wave function that appears in Eq. 
(\ref{MainEqphi}): $a$ and $\vec b$ are 
together four functions of the dimensionless
relative momentum $\vec p_{ij}$. The fourth
root in front of $\phi$ is introduced because
the measure factor, 
\begin{eqnarray}
\int {dx_2 d^2 \kappa_{24}^\perp \over x_2 x_4 }
& = &
\int {2 d^3 k_{24} \over \sqrt{m^2 +k_{24}^2}
}  \, ,
\end{eqnarray} 
needs to be symmetrized with respect to
relative momenta $\vec k_{24}$ and $\vec
k_{13}$.  The resulting potential contains
a product of $\sqrt[4]{ 1 + \vec k_{13}
^{\,2}/m^2}$ and $\sqrt[4]{ 1 + \vec k_{24}
^{\,2}/m^2}$ in denominator, and the 
integration measure becomes $d^3 k_{24}$ 
like in a non-relativistic Schr\"odinger 
equation. Thus, the leading contribution 
of the measure to $BF$ terms through $M$ 
in Eqs. (\ref{calV}) and (\ref{BFinV}) is
\begin{eqnarray}
\label{measureM}
M 
& = & 
- {1 \over 16} \, 
\left({4 \over 3} \alpha \right)^2 \,
\left( \vec p_{13} ^{\,2} + \vec p_{24}
^{\,2} \right) \, .
\end{eqnarray}

The spin contribution $1 + \alpha^2 S$ in 
Eq. (\ref{calV}) is obtained from the sum 
over quark spins
\begin{eqnarray}
\sum_{s_2 s_4} 
{j_{12}^\mu \bar j_{43 \, \mu} \over 4 m^2} \, 
{\Psi_{s_2 s_4}(\vec k_{24})  \over 
\sqrt[4]{ 1 + \vec k_{24}^{\,2}/m^2} } \, ,
\end{eqnarray}
using the two component spinors in Eq. 
(\ref{chitilde}) and the wave function in 
Eq. (\ref{phi2}). One multiplies the whole
eigenvalue equation by $\tilde \chi_1$ 
from the left and by $\sigma_3 \tilde
\chi_{\bar 3}$ from the right and sums up
over spins 1 and 3. Then, the kinetic energy
multiplies only the matrix $\phi(\vec p_{13})$,
and the potential term contains the matrix
\begin{eqnarray}
{ S_l^\mu \, \phi(\vec p_{24}) \, S_{r \mu} 
\over  
4m^2(E_{k_{13}} + m)(E_{k_{24}} + m) } \, ,
\end{eqnarray}
where, using $\alpha^\mu = \gamma^0 \gamma^\mu$
and notation from Ref. \cite{BD}, 
\begin{eqnarray}
S_l^\mu 
& = & 
{}^{\left[E_{13}+m , \vec k_{13} \vec \sigma \right]} \,
{\cal L}_{13} \,
\alpha^\mu \,
{\cal L}_{24} 
\left[\,^{E_{24} + m}_{\vec k_{24} \vec \sigma }\right] \, , \\
S_r^\mu
& = & 
{}^{\left[-\vec k_{24} \vec \sigma , E_{24} + m \right]} \,
{\cal L}_{24} \, 
\alpha^\mu \,
{\cal L}_{13} \, 
\left[\,_{E_{13} + m }^{-\vec k_{13} \vec \sigma }\right] 
\, .
\end{eqnarray}
When one neglects terms that vanish faster than $\alpha^2$
in the scaling analysis, the matrices ${\cal L}_{13}$ and 
${\cal L}_{24}$ are equivalent to 1 and the resulting 
matrix in the potential takes a fully rotationally symmetric 
form,
\begin{eqnarray}
{^{\left[E_{13}+m , \vec k_{13} \vec \sigma \right]}
\alpha^\mu 
\left[^{E_{24} + m}_{\vec k_{24} \vec \sigma }\right]  
{}^{\phi \, \left[-\vec k_{24} \vec \sigma , E_{24} + m \right]}
\alpha_\mu 
\left[_{E_{13} + m }^{-\vec k_{13} \vec \sigma }\right] 
\over  
4m^2(E_{k_{13}} + m)(E_{k_{24}} + m) } . \nonumber \\
\end{eqnarray}
The result is $\phi + (4 \alpha/3)^2 S /16$, where 
\begin{eqnarray}
S & = & 
(p_{13}^2 + p_{24}^2) \phi  
+  
\vec p_{13} \vec \sigma \,\, \vec p_{24} \vec \sigma  \,\, \phi 
+ 
\phi \,\, \vec p_{24} \vec \sigma \,\, \vec p_{13} \vec \sigma  \nonumber \\
& + &
\sigma^i \,\, \vec p_{24} \vec \sigma \,
\phi \,\, 
\vec p_{24} \vec \sigma \, \sigma^i       
+
\sigma^i \,\, \vec p_{24} \vec \sigma \,
\phi \,\, 
\sigma^i \,\, \vec p_{13} \vec \sigma \nonumber \\
& + &
\vec p_{13} \vec \sigma \, \sigma^i \,
\phi \,\, 
\vec p_{24} \vec \sigma \, \sigma^i 
+             
\vec p_{13} \vec \sigma \, \sigma^i \, 
\phi \,\, 
\sigma^i \,\, \vec p_{13} \vec \sigma  \, .
\end{eqnarray} 
The first term in $S$ is canceled by $M$ from Eq. 
(\ref{measureM}), and after summing over $i$ = 
1, 2, 3, one obtains Eq. (\ref{BFtext}).
Useful identities for Pauli matrices include
\begin{eqnarray}
\sigma^i \, \vec b \, \vec \sigma \, \sigma^i & = &  -  \vec b \, \vec \sigma \, , \\
\sigma^i \, \vec a \, \vec \sigma \, \vec b \, \vec \sigma \, \sigma^i 
& = & \vec a \, \vec \sigma \, \vec b \, \vec \sigma 
+   2 \, \vec b \, \vec \sigma \, \vec a \, \vec \sigma  \, , \\
\sigma^i \, \vec a \, \vec \sigma \, \vec b \, \vec \sigma \, \vec c \, \vec \sigma \, \sigma^i 
& = & 
- \vec a \, \vec \sigma \, \vec b \, \vec \sigma \, \vec c \, \vec \sigma 
+ 2 \, \vec b \, \vec \sigma \, \vec c \, \vec \sigma \, \vec a \, \vec \sigma \nonumber \\
& - & 
2 \, \vec a \, \vec \sigma \, \vec c \, \vec \sigma \, \vec b \, \vec \sigma \, .
\end{eqnarray}
%
\section{ Angular integrals }
%
\label{angularintegrals}
The generic form of the integrals over angles
in Eq. (\ref{MainEqphi}) is 
\begin{eqnarray}
I^{ij...l} & = &
\int d\Omega_q  {q^i q^j ... q^l \over (\vec p - \vec q)^2 } \, .
\end{eqnarray}
Using 
\begin{eqnarray}
J_n 
& = & 
\int_{-1}^1 dz {z^n \over p^2 + q^2 - 2 p q z } \, , \\
\gamma 
& = & 
{ p^2 + q^2 \over 2 p q} \, , 
\end{eqnarray}
one has
\begin{eqnarray}
J_n
& =  &
\gamma \, J_{n-1} + {(-1)^n - 1 \over 2 p q \, n } \, , 
\end{eqnarray}
and 
\begin{eqnarray}
I   
& = & 
2\pi \, J_0 \, , \\
I^i 
& = &
2\pi \, q \, J_1 \, e^i_p \, , \\ 
I^{ij}
& = &
\pi \, q^2 \, 
\left[ 
(J_0 - J_2) \, s_p^{ij}
+      
2 J_2 \, e_p^i e_p^j 
\right]  \, , \\
I^{ijk} 
& = &
\pi \, q^3 \, 
(J_1-J_3) \, 
\left(
e_p^i \, s_p^{jk} 
+ 
e_p^j \, s_p^{ik} 
+ 
e_p^k \, s_p^{ij} 
\right) \nonumber \\
& + &
\pi \, q^3 \, 
2 J_3 \, e_p^i e_p^j e_p^k  \, ,
\end{eqnarray}
where
\begin{eqnarray}
J_0 
& = & 
{1 \over pq} \, \ln{ p + q  \over
                       |p - q| } \, , \\
J_1
& =  &
\gamma \, J_0 - {1 \over p q } \, , \\
J_2
& =  &
\gamma^2 \, J_0 - {1\over p q } \, \gamma \, ,  \\
J_3
& =  &
\gamma^3  \, J_0 - {1 \over p q }\left( \gamma^2 + {1 \over 3} \right) \, ,
\end{eqnarray}
and
\begin{eqnarray}
e^i_p 
& = & 
p^i/p \, , \\ 
s^{ij}_p 
& = & 
\delta^{ij} - e_p^i e_p^j \, .
\end{eqnarray}
%
\section{ Basis functions }
%
\label{basis}
The kinetic energy and harmonic oscillator
interaction term are of the same form in 
the eigenvalue equations for all mesons,
\begin{eqnarray}
2 H_{osc} 
& = & 
\vec p^{\, 2} - k_p \, \Delta_p \, ,
\end{eqnarray}
with the spring tension $k_p$ given in Eq. (\ref{kp}).
The Hamiltonian that provides the basis for solving 
the eigenvalue equations is 
\begin{eqnarray}
2 H_b 
& = & p^2 -  k_b \, \Delta_p \, .
\end{eqnarray}
The eigenfunctions of $2H_b$,
\begin{eqnarray}
\label{phinlm}
\phi_{nlm} (\vec p) & = & \phi_{nl}(p) \, Y_{lm}(\Omega_p) \, , \\
Y_{lm}(\Omega_p) & = & \sqrt{ {2l+1 \over 4\pi} { (l-m)! \over (l+m)!} } \, 
P_l^m(\cos{\theta}) \, e^{i m \phi} \, ,
\end{eqnarray}
contain the radial wave functions $\phi_{nl}(p)$ that satisfy 
\begin{eqnarray}
\left[ p^2 - {k_b \over p^2} \partial_p p^2 \partial_p 
           + {l(l+1) k_b \over p^2 } -x_b  \right] \phi_{nl}(p)
& = & 0 .
\end{eqnarray}

In terms of the scaled variable $q = p/k_b^{1/4}$ and 
the eigenvalue $x_b = y \sqrt{k_b}$, the substitution 
$\phi(p) = \chi(q)/q$ produces
\begin{eqnarray}
-  \chi'' + {l(l+1) \over q^2 } \, \chi(q) + q^2 \, \chi  
& = & y \, \chi(q)  \, .
\end{eqnarray}
\begin{table}[t]
\caption{\label{tab:UpsilonSD} 
Coefficients $s_n$ and $d_n$ in Eq. (\ref{expansion})
in the case of the ground state of $\Upsilon$ in two 
cases corresponding to the columns third and fourth 
in Table \ref{tab:bottom}.}
\begin{ruledtabular}
\begin{tabular}{|l|r|l|r|l|}
        & fit to middle & fit to middle  &  fit to all   &  fit to all  \\
$n$     & $s_n$         & $d_n$          &  $s_n$        &  $d_n$       \\  
\hline 
1       & 0.91359      & 0.00497     &  0.73034     & 0.00532   \\
2       & 0.33678      & 0.00462     &  0.48078     & 0.00627   \\
3       & 0.17577      & 0.00403     &  0.33108     & 0.00632   \\
4       & 0.10675      & 0.00344     &  0.23735     & 0.00595   \\
5       & 0.07010      & 0.00289     &  0.17449     & 0.00539   \\
6       & 0.04807      & 0.00239     &  0.13013     & 0.00473   \\
7       & 0.03378      & 0.00195     &  0.09773     & 0.00405   \\
8       & 0.02405      & 0.00156     &  0.07356     & 0.00340   \\
9       & 0.01723      & 0.00123     &  0.05533     & 0.00280   \\
10      & 0.01237      & 0.00096     &  0.04150     & 0.00226   \\
11      & 0.00887      & 0.00074     &  0.03101     & 0.00180   \\
12      & 0.00635      & 0.00056     &  0.02308     & 0.00141   \\
13      & 0.00453      & 0.00042     &  0.01710     & 0.00110   \\
14      & 0.00322      & 0.00031     &  0.01262     & 0.00084   \\
15      & 0.00228      & 0.00023     &  0.00928     & 0.00064   
\end{tabular}
\end{ruledtabular}
\end{table}
\begin{table}[h]
\caption{\label{tab:JpsiSD} 
Coefficients $s_n$ and $d_n$ in Eq. (\ref{expansion})
in the case of $J/\psi$ in two cases corresponding to 
the columns third and fourth in Table \ref{tab:charm}.}
\begin{ruledtabular}
\begin{tabular}{|l|r|l|r|l|}
        & fit to middle & fit to middle  &  fit to all   &  fit to all  \\
$n$     & $s_n$         & $d_n$          &  $s_n$        &  $d_n$       \\  
\hline 
1       & 0.94793      & 0.01593     & 0.91858      &  0.01879  \\
2       & 0.27558      & 0.01302     & 0.33242      &  0.01631  \\
3       & 0.13007      & 0.00993     & 0.17069      &  0.01285  \\
4       & 0.07212      & 0.00723     & 0.09910      &  0.00957  \\
5       & 0.04241      & 0.00504     & 0.06014      &  0.00680  \\
6       & 0.02540      & 0.00338     & 0.03695      &  0.00463  \\
7       & 0.01524      & 0.00218     & 0.02268      &  0.00305  \\
8       & 0.00910      & 0.00137     & 0.01386      &  0.00196  \\
9       & 0.00541      & 0.00085     & 0.00843      &  0.00124  \\
10      & 0.00321      & 0.00052     & 0.00511      &  0.00078  \\
11      & 0.00190      & 0.00032     & 0.00310      &  0.00049  \\
12      & 0.00113      & 0.00019     & 0.00188      &  0.00030  \\
13      & 0.00067      & 0.00012     & 0.00114      &  0.00019  \\
14      & 0.00040      & 0.00007     & 0.00069      &  0.00012  \\
15      & 0.00024      & 0.00004     & 0.00042      &  0.00007  
\end{tabular}
\end{ruledtabular}
\end{table}
Eigensolutions normalized to $\int dq \, q^2 \, |\psi(q)|^2 
= 1$ are ($L$ denotes generalized Laguerre polynomials and 
$P(n,k)$ Pochhammer symbols)
\begin{eqnarray}
y & = & 4 n + 2 l + 3 \, , \\
\chi_{nl}(q) 
& = &
(-1)^n \sqrt{ 2 \,\, n! \over \Gamma (n + l + 3/2) } \nonumber \\
& \times &
e^{-q^2/2} \,\,  q^{l + 1} \,  L_n^{l + 1/2} (q^2) \, , \\
L_n^\lambda(x) & = &  { \Gamma (\lambda +n + 1) \over \Gamma (n+1) } \,
                   \sum_{k=0}^n { P(-n,k) \, x^k \over \Gamma (\lambda + k + 1) k! } \, , \\
P(-n,k) & = &  \Pi_{m=0}^{k-1} \, (-n+m) \, .
\end{eqnarray}
The oscillator eigenvalues and corresponding radial 
basis functions in momentum space, normalized to 1, are 
\begin{eqnarray}
x_b & = & (4 n + 2 l + 3) \, \sqrt{k_b} \, , \\
\phi_{nl}(p) & = & \chi_{nl}(p /k_b^{1/4}) \, {1 \over k_b^{1/8} \, p} \, .
\end{eqnarray}
%
\section{ Details of the wave functions }
%
\label{wf}

This appendix provides numerical data concerning the wave 
functions $S(k)/k$ and $D(k)/k$ that are shown in Figs. 
\ref{fig:swave} and \ref{fig:dwave}. Tables \ref{tab:UpsilonSD} 
and \ref{tab:JpsiSD} contain first 15 coefficients $s_n$ and $d_n$ 
in the expansion of functions $S(k)$ and $D(k)$ in the basis 
introduced in Appendix \ref{basis}, with $k_b$ = $k_p$. The 
first 15 coefficients are sufficient to see how fast the 
expansion of eigenstates in the used basis converges. The 
actual calculation of these coefficients involved 30 $s$-wave 
and 30 $d$-wave basis states. The precision with which the 
coefficients are produced in Tables VI and VII is limited to 
the digits that were stable when the number of basis states 
used in the calculation was increased above about 20 per wave.
$k_B$ is the Bohr momentum.
\begin{eqnarray}
\label{expansion}
\left[
\begin{array}{c}
S(k)  \\
D(k)
\end{array} 
\right] 
& = & 
\sum_{n=1}^\infty \,
 \left[
\begin{array}{c}
s_n \chi_{0 \, n-1}(k/k_B) \\
d_n \chi_{2 \, n-1}(k/k_B)
\end{array} 
\right]  \, . 
\end{eqnarray}
The wave functions are normalized to $\int dk (S^2 + D^2) = 1$.
\end{appendix}

\end{document}